\documentclass[reprint,doublecolumn,secnumarabic,amssymb, amsmath, aps,nofootinbib,superscriptaddress]{revtex4-1}

\usepackage{dcolumn}
\usepackage{bm}
\usepackage{graphicx}
\usepackage{color}
\usepackage{url}
\usepackage[colorlinks]{hyperref}

\newcommand{\be}{\begin{equation}}
\newcommand{\ee}{\end{equation}}
\newcommand{\bea}{\begin{eqnarray}}
\newcommand{\eea}{\end{eqnarray}}
\newcommand{\ba}{\begin{eqnarray*}}
\newcommand{\ea}{\end{eqnarray*}}
\newcommand{\barr}{\begin{array}}       
\newcommand{\earr}{\end{array}}
\def\nn{\nonumber}

\newcommand{\Sec}[1]{Sec.\,\ref{#1}}
\newcommand{\Fig}[1]{Fig.\,\ref{#1}}
\newcommand{\Tab}[1]{Table~\ref{#1}}
\newcommand{\Cit}[1]{~\cite{#1}}

\def\TeV{\ifmmode {\mathrm{\ Te\kern -0.1em V}}\else
                   \textrm{Te\kern -0.1em V}\fi}%
\def\GeV{\ifmmode {\mathrm{\ Ge\kern -0.1em V}}\else
                   \textrm{Ge\kern -0.1em V}\fi}%
\def\MeV{\ifmmode {\mathrm{\ Me\kern -0.1em V}}\else
                   \textrm{Me\kern -0.1em V}\fi}%
\def\keV{\ifmmode {\mathrm{\ ke\kern -0.1em V}}\else
                   \textrm{ke\kern -0.1em V}\fi}%
\def\eV{\ifmmode  {\mathrm{\ e\kern -0.1em V}}\else
                   \textrm{e\kern -0.1em V}\fi}%
\def\ifb{\mbox{fb$^{-1}$}}%  Inverse femtobarns.
\newcommand{\pt}      {\ensuremath{p_{\mathrm{T}}}}

\def\abseta{\ensuremath{|\eta|}}

\newcommand{\mix}{\ensuremath{|\mathcal{U}_{\mu N}|^{2}}}

\newcommand{\gev}{~\text{GeV}}

\newcommand{\ttbar}{\ensuremath{t\bar{t}}}

\begin{document}

\title{Lepton-Jets and Low-Mass Sterile Neutrinos at Hadron Colliders}

\author{Sourabh Dube}
\email{sdube@iiserpune.ac.in}
\affiliation{Indian Institute of Science Education and Research, Homi Bhabha road, Pashan, Pune 411008, India.}
\author{Divya Gadkari}
\email{divya.gadkari@students.iiserpune.ac.in}
\affiliation{Indian Institute of Science Education and Research, Homi Bhabha road, Pashan, Pune 411008, India.}
\affiliation{Department of Physics, LEPP, Cornell University, Ithaca, NY 14853, USA.}
\author{Arun M. Thalapillil}
\email{thalapillil@iiserpune.ac.in}
\affiliation{Indian Institute of Science Education and Research, Homi Bhabha road, Pashan, Pune 411008, India.}

\date{\today}

\begin{abstract}
Sterile neutrinos, if they exist, are potential harbingers for physics beyond the Standard Model. They have the capacity to shed light on our flavor sector, grand unification frameworks, dark matter sector and origins of baryon anti-baryon asymmetry. There have been a few seminal studies that have broached the subject of sterile neutrinos with low, electroweak-scale masses (i.e. $\Lambda_{\text{\tiny{QCD}}} \ll m_{N_R} \ll m_{W^\pm}$) and investigated their reach at hadron colliders using lepton jets. These preliminary studies nevertheless assume background-free scenarios after certain selection criteria which are overly optimistic and untenable in realistic situations. These lead to incorrect projections. The unique signal topology and challenging hadronic environment also make this mass-scale regime ripe for a careful investigation. With the above motivations, we attempt to perform the first systematic study of low, electroweak-scale, right-handed neutrinos at hadron colliders, in this unique signal topology. There are currently no active searches at hadron colliders for sterile neutrino states in this mass range, and we frame the study in the context of the $13\,\rm{TeV}$ high-luminosity Large Hadron Collider and the proposed FCC-hh/SppC $100\,\rm{TeV}$ $pp$-collider.
\end{abstract}

\maketitle
%------------------------------------------------------------------------------------------------------------------
\section{\label{sec:intro}Introduction}
\label{sec:1}
With the discovery of the Higgs-boson-like resonance at the LHC~\cite{{Chatrchyan:2012ufa},{Aad:2012tfa}}, we are very quickly approaching a detailed understanding of electroweak symmetry breaking and mass generation in the SM. The presence of fermion mass hierarchies (i.e. hierarchies among the Yukawa coupling constants) nevertheless remain a mystery. The Yukawa couplings that span across many orders of magnitude and the appearance of mass ratios that are seemingly very close to powers of the Cabibbo angle (see for instance~\cite{McKeen:2007ry} and references therein) along with patterns in the quark and lepton mixing matrices seem to suggest that the flavor sector of the SM may have a rich underlying structure. 

All the current experiments are largely consistent with the existence of three neutrino electroweak eigenstates $(\nu_e, \nu_\mu,\nu_\tau)$. Nevertheless, there have been a few tantalising discrepancies from various short-baseline neutrino experiments~\cite{{Aguilar:2001ty},{AguilarArevalo:2010wv},{MiniBooNE_new_antineu},{Mention:2011rk}} over the years. They have occasionally been very hard to accommodate in the three active-neutrino picture, leading to many studies incorporating additional singlet neutrino states to the framework~\cite{{Sorel:2003hf},{Nelson:2010hz},{Kopp:2011qd},{Giunti:2011ht},{Barger:2011rc},{Donini:2001xy},{Dighe:2007uf},{Donini:2007yf},Maltoni:2007zf,{Donini:2008wz},deGouvea:2008qk,{Meloni:2010zr},Kopp:2011qd,Karagiorgi:2011ut,Bhattacharya:2011ee,Giunti:2011gz}. For instance, trying to accommodate the LSND~\cite{Aguilar:2001ty} and MiniBooNE~\cite{AguilarArevalo:2010wv} anomalies with observations from solar and atmospheric neutrino measurements require $\Delta m^2_{\text{\tiny{sterile}}} \sim \mathcal{O}(1)\, \text{eV}^2$. A similar mass squared difference is also seemingly required to reconcile the reactor anti-neutrino flux deficit~\cite{Mention:2011rk}, but this interpretation has been weakened recently~\cite{An:2017osx}. On the other hand, embedding frameworks leading naturally to light neutrino masses, such as the see-saw mechanism~\cite{see-saw}, into grand unified models~\cite{Mohapatra:1974hk,Mohapatra:1974gc, Senjanovic:1975rk, Witten:1979nr, Nandi:1985uh, mohapatra:1986bd, mohapatra:1986aw, Babu:1992ia, Bajc:2006ia} furnishes singlet neutrino states that are extremely heavy with a mass $\mathcal{O}(10^{12}-10^{16})\,\rm{GeV}$. These have the added benefit of mitigating, to some extent, fine-tuning of the neutrino Yukawa coupling constants. In these models Yukawa couplings may be $\mathcal{O}(1)$ and the large hierarchy in mass is subsequently generated, after mass diagonalization. There are also intriguing models~\cite{Asaka:2005an, Asaka:2005pn,Canetti:2012vf, Canetti:2012kh} with sterile neutrino states below the $\Lambda_{\text{\tiny{QCD}}}$ scale with masses $\mathcal{O}(1)\,\rm{keV}$ that may simultaneously be able to explain structures in the lepton sector, provide dark-matter candidates as well as furnish a solution to the baryon anti-baryon asymmetry observed in the universe. Along with these considerations perhaps there is also another aspect to be kept in mind -- a small right-handed neutrino mass ($m_{\nu_R}^{\text{\tiny{M}}}$)  must be considered technically natural, as emphasised by~\cite{{Fujikawa:2004jy}, {deGouvea:2005er}}, since in the limit $m_{\nu_R}^{\text{\tiny{M}}}\rightarrow 0$ one regains $U(1)_{\text{\tiny{B-L}}}$ as a global symmetry of the Lagrangian.

The above considerations suggest that \textit{a priori} there are perhaps no immutable reasons to expect the right-handed neutrino mass-scale to be at a particular value. Motivated by this realization it is reasonable to devise search strategies for sterile neutrinos that cover all possible mass-scales. 

There has indeed been endeavours to directly and indirectly search for sterile neutrino states across various mass scales (see for instance~\cite{Keung:1983uu, Pilaftsis:1991ug, Datta:1993nm, delAguila:2006bda,Atre:2009rg, Deppisch:2015qwa,Dev:2016gvv, Antusch:2016ejd,Lindner:2016lxq,Campos:2017odj,Abada:2012mc, Abada:2013aba,Abada:2014kba} and associated references). For instance, in~\cite{Antusch:2016vyf} the sensitivity of a future lepton collider to displaced vertex searches were investigated in final states $e^+ e^- \rightarrow \nu (N \rightarrow l^\pm jj, l^+ l^- \nu,\ldots)$. A similar, earlier study~\cite{Helo:2013esa} based on displaced vertices at the LHC investigated processes such as $pp\rightarrow l^\pm (N \rightarrow l^\pm X)$. Another study in~\cite{Dib:2015oka} advocated looking for processes $W^+ \rightarrow e^+ \mu^- e^+ \nu_e$ and $W^+ \rightarrow e^+ e^+ \mu^- \bar{\nu}_\mu$ initiated by sterile neutrinos at the LHC; the latter being initiated only in the case of Majorana sterile neutrinos. Studies such as~\cite{Mitra:2016kov} focused on left-right symmetric models with a heavy $W_R$, leading to `neutrino-jet' final states $W_R\rightarrow l (N\rightarrow l jj)$; where the $N$ decay-products are collimated even for $m_N\gg m_{W^\pm}$. Recently, there have also been interesting studies attempting to constrain electroweak-scale sterile neutrinos through precision Higgs data~\cite{Das:2017zjc} and higgs decays \cite{Das:2017rsu}. A more complete discussion of current theoretical studies and limits, across various mass-scales, is contained in~\cite{delAguila:2006bda,Atre:2009rg, Deppisch:2015qwa,Dev:2016gvv, Antusch:2016ejd} and associated references. 

\begin{figure}
  \centering
   \includegraphics[width=0.5\textwidth]{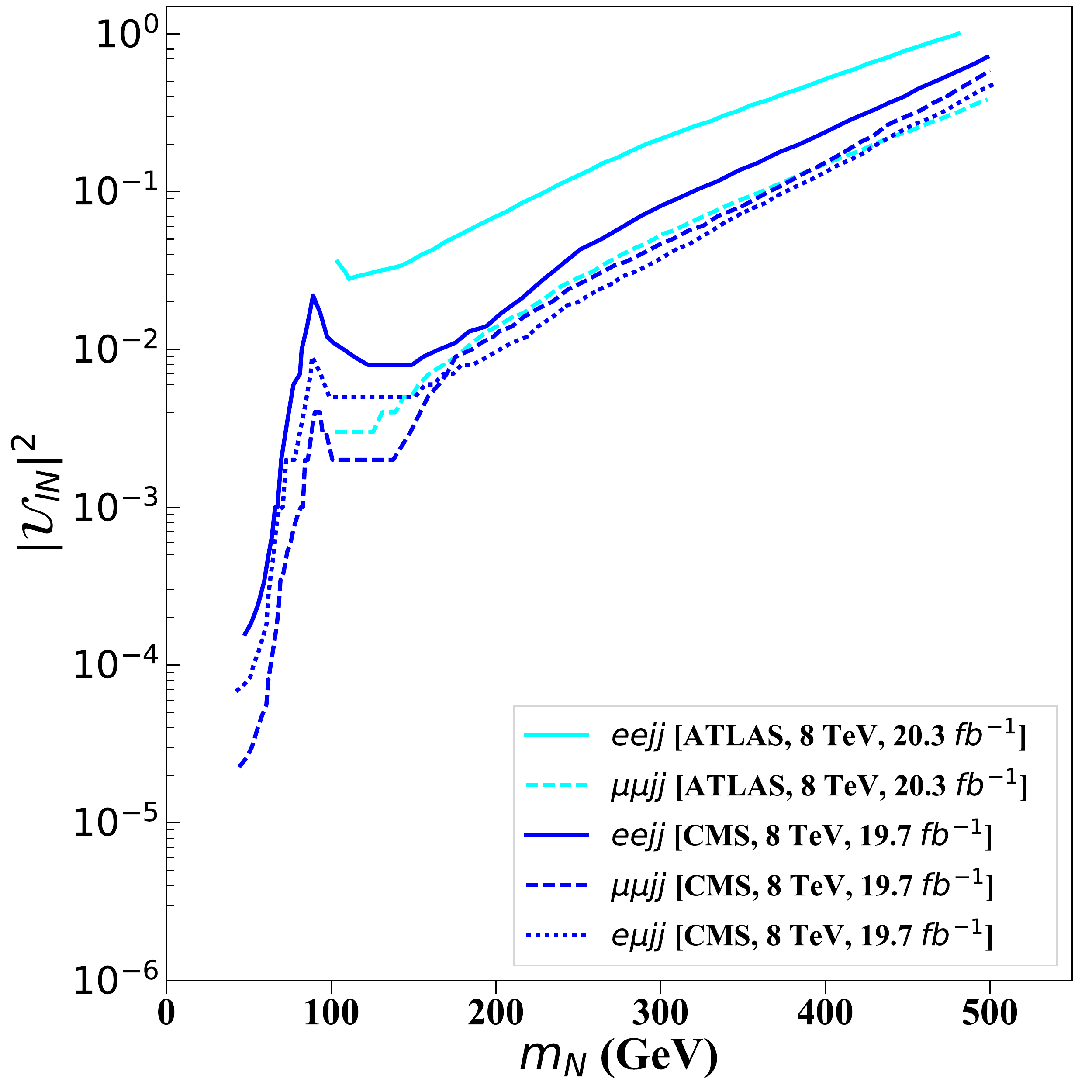}
  \caption{Current constraints on sterile-active mixing ($|\mathcal{U}_{ln}|$) from the ATLAS and CMS collaborations~\cite{Khachatryan:2015gha, Aad:2015xaa, Khachatryan:2016olu}. The relevant final states being searched for are like-sign leptons with associated jets ($l^\pm l^\pm jj$) in all the present analyses. Few of the CMS limits go all the way to intermediate masses of around $50\,\rm{GeV}$. The preliminary limit from CMS for $\sqrt{s}=13\,\rm{TeV}$ with $2.3\,\rm{fb}^{-1}$ of data~\cite{Sirunyan:2017xnz} is not shown.}
  \label{fig:hadroncolliderlim}
\end{figure}

The ATLAS and CMS collaborations have performed dedicated searches for heavy Majorana neutrinos~\cite{Khachatryan:2015gha, Aad:2015xaa, Khachatryan:2016olu,Sirunyan:2017xnz} in various channels. The CMS collaboration has looked for heavy sterile neutrinos in $\mu^\pm \mu^\pm jj$, $e^\pm e^\pm jj$ and $e^\pm\mu^\pm jj$ final states at $\sqrt{s}=8\,\rm{TeV}$ with  $19.7\,\rm{fb}^{-1}$ of data~\cite{Khachatryan:2015gha,Khachatryan:2016olu}. The ATLAS collaboration has similarly searched for heavy Majorana neutrinos in the $\mu^\pm \mu^\pm jj$ and $e^\pm e^\pm jj$ channels at $\sqrt{s}=8\,\rm{TeV}$ using $20.3\,\rm{fb}^{-1}$ of collected data~\cite{Aad:2015xaa}. The CMS collaboration has also recently set preliminary limits at $\sqrt{s}=13\,\rm{TeV}$~\cite{Sirunyan:2017xnz}, with $2.3\,\rm{fb}^{-1}$ data, for heavy composite Majorana neutrinos in final states with two leptons and two quarks. All the current LHC constraints for the $l^\pm l^\pm jj$ channels are summarised in \Fig{fig:hadroncolliderlim}.

We are interested in probing a regime where the sterile neutrino states have a mass above the bottom-quark mass ($m_b$) but is at the same time well below $m_{W^\pm}$ 
\ba
\Lambda_{\text{\tiny{QCD}}} \ll m_b < m_{N_R} \ll m_{W^\pm} \; .
\ea

\begin{figure}
  \centering
   \includegraphics[width=0.45\textwidth]{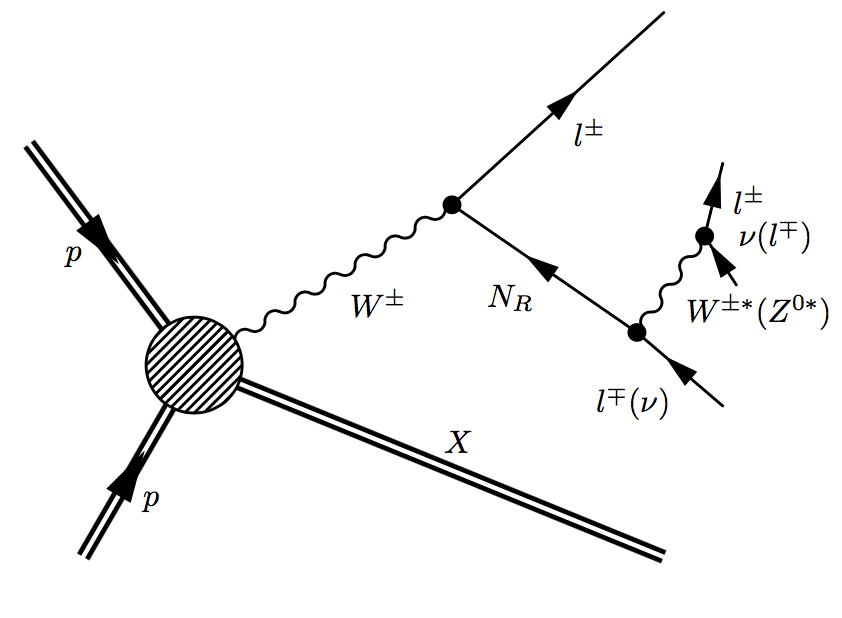}
  \caption{In the mass regime of interest, $m_b < m_{N_R} \ll m_{W^\pm}$, the main production channel at hadron colliders is through single-$W^\pm$ production and decay. Since $m_{N_R}/m_{W^\pm} \ll 1$ the leptons from the $N_R$ decay are collimated and form a displaced lepton jet in the relevant parameter space~\cite{Izaguirre:2015pga}. The lepton from the initial $W^\pm$ decay is detected as a prompt lepton.}
  \label{fig:signal_event}
\end{figure}

In this narrow mass-region, the existing constraints are minimal and the signal topology is unique while being challenging. We shall sharpen and motivate the region of interest in more detail in \Sec{sec:3}. The prototypical signal event is illustrated in \Fig{fig:signal_event}. In this region the sterile neutrino is usually very boosted and the decay products get collimated into a lepton jet.

\be
pp \rightarrow l^\pm + (N_R \rightarrow \text{Lepton Jet})+X
\ee

There have been a few hadron collider studies specifically focused on this region~\cite{Izaguirre:2015pga, Antusch:2016ejd}. The pioneering study \cite{Izaguirre:2015pga} assumed a background-free search, employing certain selection criteria, with cosmic-ray initiated muon bundles estimated based on an ATLAS analysis~\cite{Aad:2014yea}; the latter looked for long-lived neutral particles in LHC events with two lepton jets. Based on these estimates, limits at $13\,\rm{TeV}$ LHC are set in this region, assuming an integrated luminosity of $300\,\rm{fb}^{-1}$. Similarly, the study pertaining to this mass-scale discussed in~\cite{Antusch:2016ejd} for $pp$-colliders, assumes that there are no backgrounds for $1\,\rm{mm}<c\tau<1\,\rm{m}$ vertex displacements. With this and a few other assumptions, the study estimates preliminary limits for the high-luminosity LHC (HL-LHC) at $13\,\rm{TeV}$ and the FCC-hh/SppC $pp$-collider at $100\,\rm{TeV}$~\cite{Golling:2016gvc,CEPC-SPPCStudyGroup:2015csa}. They conclude by acknowledging that a realistic estimate of the backgrounds and sensitivities is very much required in this mass-regime.

As we shall discuss in Secs.\,\ref{sec:3} and \,\ref{sec:4}, the above search methodologies, selection criteria and considerations regarding signal and backgrounds have to be drastically modified under realistic conditions. Our aim is to perform a systematic study in this mass-scale regime and investigate realistic selection criteria that optimise searches for these light sterile neutrinos at hadron colliders. Towards this aim we explore the discovery potential at the $13\,\rm{TeV}$ LHC and the proposed FCC-hh/SppC $100\,\rm{TeV}$ $pp$-collider~\cite{Golling:2016gvc,CEPC-SPPCStudyGroup:2015csa}. In this low mass-scale regime the decay products from the right-handed neutrino get collimated into a narrow cone~\cite{Izaguirre:2015pga}. As we elaborate in Secs.\,\ref{sec:3} and \ref{sec:4}, we will therefore optimise for a topology consisting of a prompt lepton and a collimated set of muons, a muon lepton-jet. 

One of the main constraints in the region of interest comes from electroweak precision data~\cite{delAguila:2008pw, Akhmedov:2013hec, Basso:2013jka, Blas:2013ana, Antusch:2015mia}. To very good approximation, the limits on active-sterile mixing ($|\mathcal{U}_{lN}|^2$), from electroweak precision data, are found to be almost independent of the sterile neutrino masses in this region. At 90\% confidence level they are approximately given by $|\mathcal{U}_{e N}|^2 \leq 3\times 10^{-4}\, , |\mathcal{U}_{\mu N}|^2 \leq 1.5\times10^{-4}$ and  $|\mathcal{U}_{\tau N}|^2 \leq 13\times10^{-4}$~\cite{Deppisch:2015qwa, delAguila:2008pw, Akhmedov:2013hec, Basso:2013jka, Blas:2013ana, Antusch:2015mia}.

The other major constraint in this mass regime comes from limits on heavy sterile states produced in $Z^0$ decays. The L3~\cite{Adriani:1992pq} and DELPHI~\cite{Abreu:1996pa} collaborations have performed a reanalysis of the LEP data in this context. The former sets a limit $|\mathcal{U}_{lN}|^2\lesssim (0.7-1.0)\times 10^{-4}$, corresponding to a limit on the branching ratio $\text{Br}(Z^0\rightarrow \bar{\nu} N)\lesssim 10^{-5}$ ~\cite{Shevchenko:1993fe}, in the region of interest. The DELPHI analysis puts a limit $\text{Br}(Z^0\rightarrow \bar{\nu} N)\lesssim 1.3\times10^{-6}$ at 95\% C.L. which corresponded to $|\mathcal{U}_{lN}|^2\lesssim 10^{-5}$ ~\cite{Abreu:1996pa}.

In \Sec{sec:2}, to clarify notations and put our study in context, we briefly discuss the well known theoretical motivations for sterile neutrinos. Here, we also briefly consider models where low-mass right-handed neutrinos could arise in a natural way. In \Sec{sec:3} we then discuss the unique signal topology furnished by sterile neutrinos in the mass regime of interest and also discuss aspects of the various relevant backgrounds. Then, in \Sec{sec:4} we present our analysis methodologies and main results. We summarise our pertinent findings in \Sec{sec:5}.

%------------------------------------------------------------------------------------------------------------------
\section{\label{sec:2} Right-Handed Sterile Neutrinos and the Standard Model}
\label{sec:2}

The inexplicable and large hierarchies among the fermion masses manifests in its most extreme form in the case of neutrinos. To clarify notations and set context we briefly consider the theoretical underpinnings behind sterile neutrinos and specific models where  low, electroweak-scale masses could be generated for these states.

Neutrino oscillation experiments only furnish information about mass-squared differences\Cit{Olive:2016xmw}. Through careful endpoint measurements of the tritium $\beta$-decay spectrum, Troitzk~\cite{troitsk} and Mainz~\cite{mainz} experiments were able to put an upper limit at $95\%$ C.L. of about
\bea
m_\nu < 2~\text{eV}\; .
\eea

Light neutrinos play a significant role in cosmology, by effecting the expansion history and the growth of primordial structures, which in combination with other astrophysical and cosmological observations, lead to an even tighter bound~\cite{Reid:2009nq,GonzalezGarcia:2010un,Ade:2015xua}
\ba
\sum m_\nu < 0.23~\text{eV}\; .
\ea

The KATRIN experiment~\cite{KATRIN} is expected to reach a sensitivity close to $m_{\nu}<0.2\,\text{eV}$ as well. 

All these observations suggest that the neutrinos in the SM have a mass-scale in the sub-eV regime. The neutrinos with SM quantum numbers thus seem to have a mass-scale at least a million times smaller than the next heaviest fermion, the electron. If not an accident of nature, these small neutrino masses beg for an explanation.

In the standard model, the neutrinos have just a single left-helicity field associated with them and therefore one cannot directly write a Dirac mass term in the usual way. One could of course extend the framework minimally by adding just a right-handed helicity neutrino field, thereby giving neutrinos a Dirac mass after electroweak symmetry breaking\footnote{Flavor indices are suppressed in the following discussions for clarity.}
\be
\mathcal{L}_{\text{\tiny{mass}}} \supset   m_\nu^{\text{\tiny{D}}}\,\left( \bar{\nu}_L \nu_R + \bar{\nu}_R \nu_L \right)    \equiv  m_\nu^{\text{\tiny{D}}}\,\bar{\nu} \nu  \; .
\ee

If this was the only contribution, the relevant yukawa coupling here has to be fine-tuned to a very small value, to be consistent with the sub-eV mass-scale of the neutrinos. The interesting observation is that, since the right-handed neutrino field carries no SM charges, one is allowed to also write an additional contribution to the mass of the form 
\be
\mathcal{L}_{\text{\tiny{mass}}} \supset m_{\nu_R}^{\text{\tiny{M}}}  \left(  \bar{\nu}^{\text{\tiny{c}}}_R \nu_R + \bar{\nu}_R \nu_R^{\text{\tiny{c}}} \right) \equiv m_{\nu_R}^{\text{\tiny{M}}} \,\bar{\chi} \chi \; ,
\ee
i.e. a Majorana mass term. Here, the charge conjugation is defined as $\psi^{\text{\tiny{c}}}=i \gamma^2 \psi^*$, with the notation $\psi^{\text{\tiny{c}}}_R=(\psi_R)^{\text{\tiny{c}}}$ and the Majorana field $\chi$ is defined to be $\chi=\nu_R+\nu_R^{\text{\tiny{c}}}$. Note that a similar term with $\nu_L$ would be forbidden in this minimal scheme due to SM gauge invariance -- the $\nu_L$ field is part of the $SU(2)_L$ doublet with non-zero hypercharge.

This additional contribution enables a novel way in which the very-small neutrino masses could be generated -- the so called see-saw mechanism~\cite{see-saw}. As motivated above, in its simplest form it leads to a neutrino mass-matrix of the form
\be
 \mathcal{M}_\nu =  
\left ( 
\begin{tabular}{c|c}
0 &  $ \frac{1}{2} m_\nu^{\text{\tiny{D}}}$\\
\hline
 $ \frac{1}{2} m_\nu^{\text{\tiny{D}}~\text{\tiny{T}}}$ & $m_{\nu_R}^{\text{\tiny{M}}} $ \\
 \end{tabular}
 \right)\;.
\ee
Taking for example the simplest 1-flavor case, with a $2\times2$ mass-matrix, leads to mass-eigenvalues
\bea
m_{1,2}=\frac{1}{2}\left[m_{\nu_R}^{\text{\tiny{M}}} \pm \sqrt{m_{\nu_R}^{\text{\tiny{M}}\,2} + m_\nu^{\text{\tiny{D}}\,2} } \right]\; ,
\eea
with two Majorana eigenstates
\bea
\nu_1 &=& \tilde{\chi} \cos\theta - \chi \sin\theta \\ \nn
\nu_2 &=& \tilde{\chi} \sin\theta + \chi \cos\theta \; .
\eea
Here, $\chi=\nu_R+\nu_R^{\text{\tiny{c}}}$ as before, $\tilde{\chi}=\nu_L+\nu_L^{\text{\tiny{c}}}$ and the mixing angle is defined as
\be
\tan2\theta= - \frac{m_\nu^{\text{\tiny{D}}}}{m_{\nu_R}^{\text{\tiny{M}}}}\; .
\ee

If one assumes that $m_{\nu_R}^{\text{\tiny{M}}} \gg m_\nu^{\text{\tiny{D}}}$, then one obtains a light and heavy neutrino state, as is well known, 
\bea
\nu_l \sim \tilde{\chi} \; , ~\nu_h \sim \chi \; , 
\eea
with masses
\bea
m_l \sim - \frac{m_\nu^{\text{\tiny{D}}\,2}}{m_{\nu_R}^{\text{\tiny{M}}}}\; , ~m_h \sim m_{\nu_R}^{\text{\tiny{M}}}\; .
\eea
Observe that the heavier state is a right-handed majorana fermion.

Note also from the above discussions that the mixing matrix elements $|\mathcal{U}_{lN}|$, between  active and right-handed (sterile) states, roughly scale like $m_\nu^{\text{\tiny{D}}} m_{\nu_R}^{\text{\tiny{M}}\,-1}$. If one had $m_\nu^{\text{\tiny{D}}} \sim \mathcal{O}(\text{EW-scale})$, hence seemingly mitigating to some extent the relative hierarchy among Yukawa couplings, then this would imply $m_{\nu_R}^{\text{\tiny{M}}} \sim \mathcal{O}(10^{12}-10^{15}\,\rm{GeV})$ to get viable light neutrino masses in this simplest framework. This right-handed Majorana scale is also attractive from the point of view grand unified theories\Cit{Langacker:1980js}, specifically left-right symmetric grand unified models such as the Pati-Salam model\Cit{Pati:1974yy}. The above discussions may be extended to the case of two or more sterile neutrinos. The inclusion of additional sterile neutrinos to the three active ones adds more structure to the neutrino sector.

On the other hand, as we alluded to before, it must be noted that low $m_{\nu_R}^{\text{\tiny{M}}}$ scales must be considered technically natural~\cite{{Fujikawa:2004jy}, {deGouvea:2005er}} -- since in the limit $m_{\nu_R}^{\text{\tiny{M}}}\rightarrow 0$ one regains $U(1)_{\text{\tiny{B-L}}}$ as a global symmetry of the Lagrangian. In this context, the presence of additional states in a k-neutrino framework furnishes new possibilities. One could now have novel flavor structures, under seesaw or non-seesaw scenarios, sometimes augmented by lepton-number-like family symmetries. In many of these models the right-handed neutrino mass-scale is unconstrained and could in general be small, leading to interesting observational consequences~\cite{{Fujikawa:2004jy},{deGouvea:2005er},Pilaftsis:1991ug,{Wyler:1982dd},{Mohapatra:1986aw},{Mohapatra:1986bd},{Langacker:1998ut},{Chacko:2003dt},{Gherghetta:2003hf},{Malinsky:2005bi},{Shaposhnikov:2006nn},{deGouvea:2006gz},{Kersten:2007vk},{Gavela:2009cd},{deGouvea:2011zz}}. 

All of these thus imply, as we mentioned earlier, that \textit{a priori} it is prudent to be agnostic about the exact $m_{\nu_R}^{\text{\tiny{M}}}$ scale and devise search strategies that would span the full range of possibilities. We will be specifically interested in scenarios where $m_{\nu_R}^{\text{\tiny{M}}}\ll m_{W^\pm}$, i.e. in the low, electroweak-scale regime. In these scenarios the mixing between the active-sterile states could be larger than naive expectations and potentially unsuppressed. For instance, in inverse seesaw models~\cite{{Mohapatra:1986aw},{Mohapatra:1986bd}} -- the simplest realization of which has three extra standard model singlet neutral fermions ($\Psi$) in addition to three generations of sterile ($N_R$) and active neutrinos ($\nu_L$)-- the Lagrangian takes the form
\bea
\mathcal{L}^{\text{\tiny{inv. see-saw}}}_{\text{\tiny{mass}}} \supset -M_{\text{\tiny{D}}}\bar{\nu}_L N_R-M \bar{\Psi}_L N_R-\frac{\delta}{2} \bar{\Psi}_L \Psi_L^{\text{\tiny{c}}} \; , ~~~
\eea
which leads to a neutrino mass-matrix 
\be
 \mathcal{M}_\nu =  
\left ( 
\begin{tabular}{ccc}
0 & $M_{\text{\tiny{D}}}$ &0 \\
  $M_{\text{\tiny{D}}}^{\,\text{\tiny{T}}}$&0 & $M$ \\
 0&$M^{\,\text{\tiny{T}}}$ & $\delta$
 \end{tabular}
 \right)\;.
\ee

Note that as $\delta\rightarrow 0$ one regains the lepton-number-like protection symmetry, and hence a small $\delta$ is technically natural. 

On diagonalizing the mass matrix, assuming a hierarchy among the scales $\delta \ll M_D \lesssim M$, the mass of the light neutrinos scale as 
\be
m_\nu \sim \delta\,\frac{M_{\text{\tiny{D}}}^2}{M^2}\; ,
\ee
while the active-sterile mixing matrix elements still scale as
\be
|\mathcal{U}_{lN}| \sim \frac{M_{\text{\tiny{D}}}}{M}\; .
\ee

Owing to the presumably small $\delta$ and the difference in scaling behavior between masses and mixing angles, we have the possibility of getting very small SM neutrino masses, while retaining the possibility of relatively light sterile neutrinos. The latter could also have significant mixing with active neutrinos ($|\mathcal{U}_{lN}| \sim M_{\text{\tiny{D}}}/M \lesssim \mathcal{O}(1)$). This could lead to effective couplings between the sterile states and the $W^\pm$, $Z^0$ vector gauge bosons that are relatively unsuppressed. Thus, the relatively large mixing angles along with the lighter masses open the way for these right-handed sterile states to be searched for in particle collider experiments.

In the mass-range we are interested in, $m_b < m_{N_R} \ll m_{W^\pm}$, the dominant production mode for a sterile neutrino are through $W^\pm$ charged-current and $Z^0$ neutral-current interactions, mediated through the mixings with active-neutrino states, 
\ba
pp &\rightarrow& W^\pm+X \rightarrow l^\pm N_R+X \; ,\\ 
&\hookrightarrow& Z^0+X \rightarrow \nu N_R+X \; .
\ea
For higher $m_{N_R}$ and energies, other production modes also become relevant\Cit{Dev:2013wba,Alva:2014gxa,Hessler:2014ssa,Degrande:2016aje,Ruiz:2017yyf}. For masses below the bottom-quark mass, $m_{N_R}<m_b$, production through $B$-meson decay channels also open up.

The right-handed neutrinos after being produced subsequently decay, again through $W^\pm$ charged-current or $Z^0$ neutral-current interactions mediated by active-sterile mixing. The partial widths to leptonic final states are given by~\cite{{Atre:2009rg},{Helo:2010cw}}
\bea
&\Gamma(N_R \xrightarrow{W^*} l_a^- l_b^+ \nu_b)~~=& \frac{G_F^2}{192 \pi^3} m_N^5 |\mathcal{U}_{aN}|^2\; ,\\ \nn
&\Gamma(N_R \xrightarrow{W^*/Z^*} l_a^- l_a^+ \nu_a) =& \frac{G_F^2}{96 \pi^3} m_N^5 |\mathcal{U}_{aN}|^2\; , \\ \nn
&~~~~~~~~&\left( \hat{g}_L^2+\hat{g}_L \hat{g}_R+\hat{g}_R^2 +2 \hat{g}_L+ 2 \hat{g}_R+1 \right) \; , \\ \nn
&\Gamma(N_R \xrightarrow{Z^*} \nu_a l_b^+ l_b^-)~~=& \frac{G_F^2}{96 \pi^3} m_N^5 |\mathcal{U}_{aN}|^2 \left(\hat{g}_L^2+\hat{g}_L \hat{g}_R+\hat{g}_R^2\right) \; ,\\ \nn
&\Gamma(N_R \xrightarrow{Z^*} \nu_a \nu_b \bar{\nu}_b)~~=& \frac{G_F^2}{768 \pi^3} m_N^5 |\mathcal{U}_{aN}|^2 \; .
\label{eq:N_decayrate}
\eea
Here the SM couplings are defined as $\hat{g}_L=\frac{1}{2}+\sin^2\theta_{\text{\tiny{W}}}$ and $\hat{g}_R=\sin^2\theta_{\text{\tiny{W}}}$.

We are interested in devising an optimal search strategy for low, electroweak-scale sterile neutrinos produced at $pp$-colliders; dominantly via decays of $W^\pm$ and which decay through their charge-current interactions. The typical process of interest is therefore
\be
pp \rightarrow W^\pm+X \rightarrow l_a^\pm (N_R\rightarrow l_a^\pm l_a^\mp \nu_a)+X  \;,
\ee
as shown in \Fig{fig:signal_event}.

In the next section we will take a closer look at the event and background toplogies to be expected and discuss considerations that must be taken into account for an effective search at the LHC and the proposed $100\,\rm{TeV}$ $pp$-colliders.

%------------------------------------------------------------------------------------------------------------------
\section { \label{sec:3} Event Topology and Backgrounds}
\label{sec:3}
Traditional multilepton searches (for example Ref.~\cite{CMS:2017wua, Aad:2014hja})  can have high sensitivity for sterile neutrinos with masses above $\sim$100\gev.
These searches rely on prompt, well-separated (and isolated) leptons in the final state, and typically require lepton transverse momenta (\pt) to satisfy
$\pt > 20\gev$. With careful selection of isolation criteria, and lowering of the lepton \pt\ threshold, or by considering final states with dileptons
and jets, the sensitivity can be extended to sterile neutrino masses as low as 50\gev~\cite{Khachatryan:2015gha, Khachatryan:2016olu}. 

However as the mass of the sterile neutrino becomes lighter, which is the case of interest in our current investigation, new search strategies need to be explored. One such
interesting final state involves a prompt lepton along with a lepton-jet~\cite{Izaguirre:2015pga}. In a lepton-jet, two or more leptons lie very close to each other in the detector. Such a signature will be rejected by standard isolation criteria, and needs a separate, special selection criteria. Lepton-jet 
searches have been carried out by the LHC experiments in the context of different new physics models. ATLAS searches for pairs of lepton-jets~\cite{ATLAS:2016jza} which might or might not be significantly displaced from the interaction
point. CMS searches for a pair of leptons~\cite{CMS:2014hka} which may lie close to each other, but which are displaced and need to satisfy $m_{\ell\ell}>15\gev$. The CMS search also has stiff requirements on lepton \pt.

A different approach is needed to probe for sterile neutrinos that lie in the range $m_b < m_{N_R} \ll m_{W^\pm}$. We choose a final state that
consists of a lepton-jet accompanied by the presence of a prompt, well-isolated lepton. This choice of final state dictates the analysis strategy, since it significantly affects which standard model
processes will act as a background to the search. Further selections to optimize the sensitivity are then governed by the interplay between the signal of interest, the backgrounds
and the specific selections.

The particular decay chain we probe is  $W^\pm \rightarrow \ell^{\pm} N_R \rightarrow \ell^{\pm} \ell'^{\pm}\ell''^{\mp}\nu$. Here the $\ell$
arises promptly from the decay of the $W^\pm$, and has relatively high \pt. On the other hand, the $\ell'$ and $\ell''$, which arise from the decay
of the sterile neutrino ($N_R$), are not necessarily prompt and can have low \pt, depending on the mass and lifetime of the $N_R$. Moreover, depending
on the boost of the $N_R$, the $\ell'$ and $\ell''$ can also come close to each other forming a lepton-jet. Let us now consider each aspect of this signal topology carefully.

The separation between the decay products of $N$ scale as $\Delta R \sim m_{N_R}/p_T^{N_R}$ while in the rest frame of the $W^\pm$ the momentum of the sterile neutrino scales as $p_T^{N_R}\sim(m_{W^\pm}^2-m_{N_R}^2)/m_{W^\pm}$. This implies that around $m_{N_R} \sim 20\gev$ the opening angle for the decay products will exceed $\Delta R \sim 0.5$. At around this mass the lepton-jet selection criteria will therefore become less efficient and one expects that the limits obtained near $m_{N_R} \sim 20\gev$ should be weaker than those from other experiments~\cite{delAguila:2008pw, Akhmedov:2013hec, Basso:2013jka, Blas:2013ana, Antusch:2015mia, Adriani:1992pq, Abreu:1996pa}. On the other hand, below $m_{N_R} \lesssim 4\gev$ there are already very strong limits on active-sterile mixing angles from other searches -- lepton number violating meson decays, peak searches in meson decays, beam dump experiments and so on (Please see for instance~\cite{Deppisch:2015qwa} and references therein). Some of the planned experiments in this mass range, such as DUNE~\cite{Acciarri:2016crz,Acciarri:2015uup} and SHiP~\cite{Alekhin:2015byh, DeLellis:2017rfg}, are projected to have the capability to probe mixing angles all the way down to $\mix \sim 10^{-10}$. We therefore sharpen our mass-regime of interest to be between
\be
4\gev ~<m_{N_R}<~25\gev\; .
\ee

In this mass regime of interest, for small sterile-active mixing $|\mathcal{U}_{\mu N}|^2$, one also expects from Eq.\,(\ref{eq:N_decayrate}), the lepton-jet to be displaced appreciably from the primary vertex due to the large $N_R$-boost. This displacement may potentially be leveraged to discriminate between signal and background. Nevertheless, as we shall explain in \Sec{sec:4} this criterion turns out to be less significant, rather counterintuitively from naive reasoning, for overall signal sensitivity.

\begin{figure}
  \centering
   \includegraphics[width=0.45\textwidth]{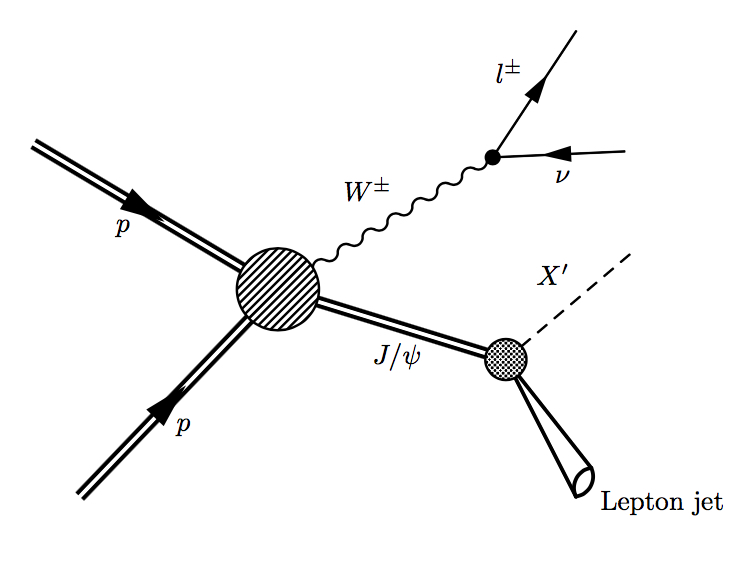}
  \caption{A possible background from heavy flavor decays. Here, a $J/\psi$ resonance is produced in association with a $W^\pm$. The decay products from the boosted $J/\psi$ fake a lepton-jet while the $W^\pm$ furnishes a prompt lepton. A relatively significant fraction of such events could still survive after a naive selection. They must therefore be accounted for more carefully while making an optimised analysis.}
  \label{fig:wcc_event}
\end{figure}

The presence of the prompt isolated lepton in the signal topology significantly simplifies the trigger needed for such a topology. Typical isolated lepton triggers at the ATLAS or CMS experiments  have $\pt$ thresholds ranging from
23\gev\ for muons to 35\gev\ for electrons. In addition, advanced trigger strategies such as those employed in Ref.~\cite{ATLAS:2016jza} can also be constructed.
At a hadron collider such as the LHC, a large source of single isolated prompt leptons is SM $W^\pm$ production.
Along with direct production, $W^\pm$ also arise through the decay of $t$-quarks. The cross sections for $t$-quark production ($\ttbar$ and single-top events) are also significant compared to signal at the LHC. Other standard model processes that give rise to more than one isolated prompt leptons ($Z/\gamma*$, or $WZ$ production) can also lead to background to a final state with a prompt lepton.

Muons generally provide cleaner lepton and lepton-jet signals as compared to electrons or $\tau$ leptons. Muons
are reconstructed using the tracking chambers and therefore gives a better lepton-jet discriminant. For our conservative estimates we shall therefore assume that there is only appreciable mixing between a single sterile state and the muon-neutrino ($\nu_\mu$). After analysis, this would therefore translate to a stricter limit on $|\mathcal{U}_{\mu N}|^2$ as a function of the sterile neutrino mass $m_{N_R}$. If other channels are open then the limits obtained in a prompt-muon and muon-lepton-jet final state analysis will be weaker.

The requirement of a lepton-jet should significantly reduce the $W^\pm$ and top backgrounds. But at hadron colliders, $W^\pm$ are typically accompanied by light hadrons in a large fraction of events. Several light hadrons such as the $J/\psi$ and the $\Upsilon$ decay to a pair of oppositely-charged leptons. When these light hadrons are boosted, the resultant dilepton decay may mimic the lepton-jet of the signal (\Fig{fig:wcc_event}). This
background can be reduced by raising the $\pt$ thresholds on the lepton-jet muons and placing strict requirements on hadronic activity. However given that almost 6\% of all $J/\psi$'s decay to a purely dimuon final state, such
requirements will not remove this background completely.

\begin{figure}
  \centering
   \includegraphics[width=0.45\textwidth]{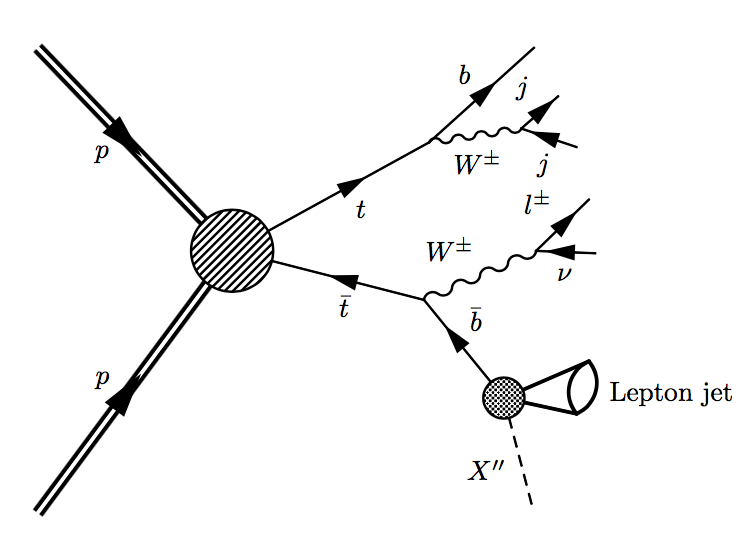}
  \caption{A prototypical background topology that may arise from $t\bar{t}$ events. Even vetoing for hadronic activity and imposing isolation requirements, potentially a large fraction of such events could contaminate the signal region.}
  \label{fig:tt_event}
\end{figure}

The other significant background could come from $\ttbar$ events. The $\ttbar$ semi-leptonic decay chain results in one prompt lepton, two $b$-quarks,
and two light quarks ($\ttbar\rightarrow W^\pm bW^\mp \bar{b}\rightarrow \ell^\pm \nu q\bar{q}' b\bar{b}$). A potential decay chain
for the $b$-hadrons is through semileptonic decay to $c$-hadrons which subsequently decay semileptonically to lighter particles. Such a decay chain can also give rise to two oppositely-charged leptons. Given the boost of the $b$-quark, these two leptons can mimic the signature of a lepton-jet.
Thus both $W$-boson production, and $\ttbar$ production can result in significant background to a prompt lepton $+$ lepton-jet final state. An example of a final state that could arise from the $\ttbar$ background is illustrated in \Fig{fig:tt_event}. A requirement of low hadronic activity in the event will supress the $\ttbar$ background considerably, but not remove it completely. Further supression can be obtained by requiring the lepton-jet to be isolated, i.e. by requiring
low hadronic activity in the immediate neighbourhood of the lepton-jet.

Other small contributions arise from single-top production, as well as low rate processes such as $Z\rightarrow 4\ell$, $Z+b\bar{b}$, or $WZ/W\gamma^*$. The $Z\rightarrow 4\ell$ process where an asymmetric internal conversion
takes place~\cite{Gray:2011us} can result in a soft muon appearing almost collinear to one of the muons from the $Z$-decay. This background as well as background from $Z+b\bar{b}$ and $WZ$ can be reduced to
negligible levels by vetoing events that have more than one isolated lepton, and by requring that the invariant mass of all three muons in the event is below the $W$-mass. The $W\gamma^*$
background can be further reduced by considering the alignment of the missing energy with the $\mu\text{-Jet}$.

\begin{figure}
  \centering
   \includegraphics[width=0.5\textwidth]{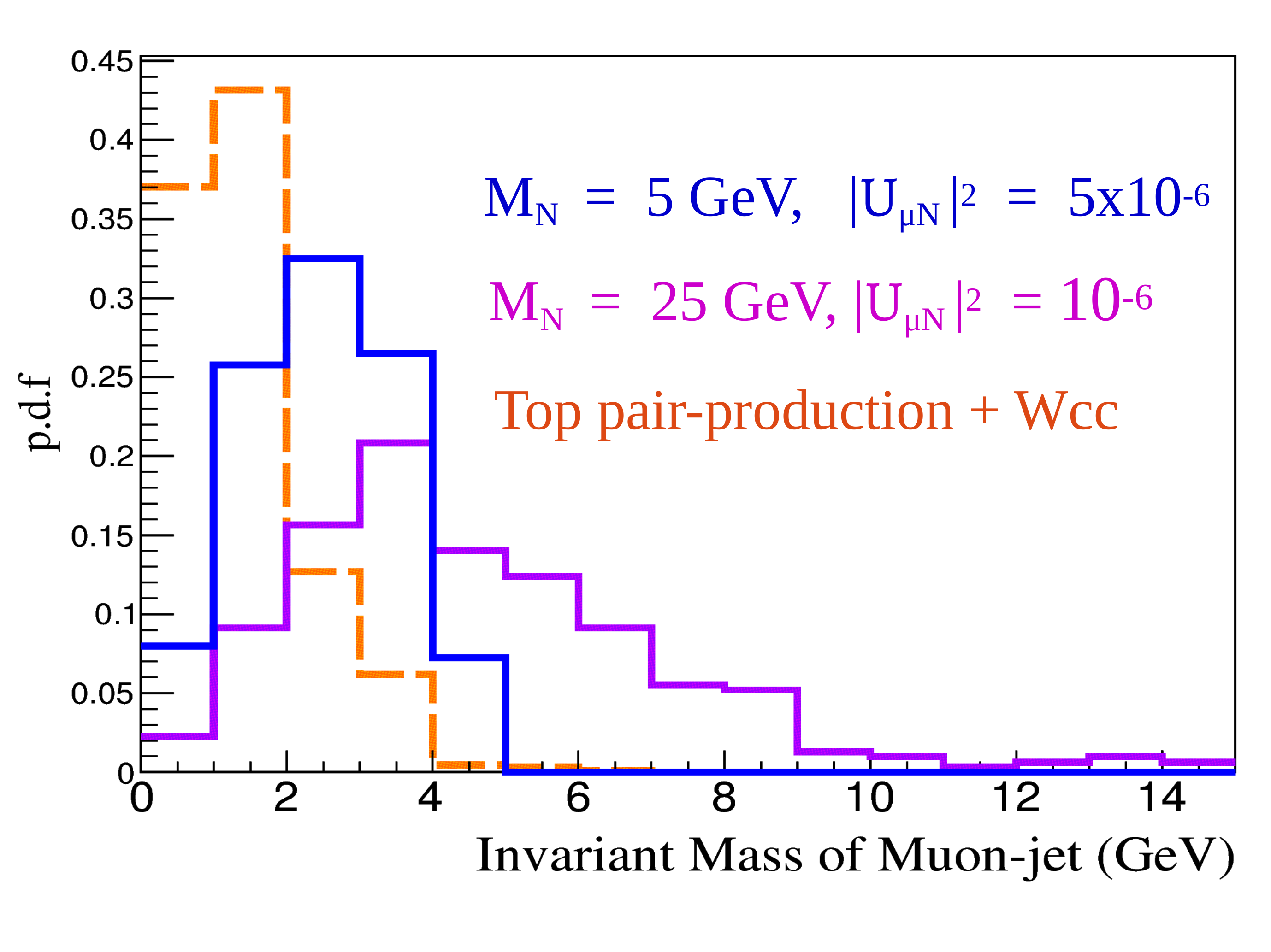}
  \caption{The invariant mass of the $\mu\text{-Jet}$ muons is shown for two signal points and for the combined background ($\ttbar$ and $W+$jets) for a 13~\TeV collider. }
  \label{fig:invmass_mj}
\end{figure}

\Fig{fig:invmass_mj} shows the invariant mass constructed from the two muons that form the $\mu\text{-Jet}$. The signal distribution, as expected, peaks at harder values
with increasing $m_{N_R}$. The background is concentrated at low values, since it arises primarily from $b$-hadron as well as lighter hadron decay.
We do not use the invariant mass in our study as we find that it inordinately affects signal acceptance for low mass sterile neutrinos.
We now proceed to detail the search strategy and discuss the prospective reach attainable at hadron colliders.

%------------------------------------------------------------------------------------------------------------------
\section{\label{sec:4} Lepton Jet probes of Sterile Neutrinos at the LHC and FCC-\lowercase{hh}/S\lowercase{pp}C }
\label{sec:4}

The signal mass-region of interest presents unique challenges and necessitates a careful analysis strategy, taking into consideration all the features of the signal and background topologies discussed in the previous section. Our aim is to carefully account for the relevant backgrounds and tailor the selection criteria to enable an optimal search strategy at a hadron machine. We will present our results in the context of the $13\,\TeV$ HL-LHC, and the proposed $100\,\TeV$ FCC-hh/SppC colliders~\cite{Golling:2016gvc,CEPC-SPPCStudyGroup:2015csa}. 

As mentioned, we will assume that there is appreciable mixing only between the sterile state and muon-neutrinos to set a conservative limit. This is partially motivated by the fact that at the LHC, muons will provide a cleaner lepton and lepton-jet signal as compared to electron or tau leptons. Identifying lepton-jets with electrons and taus require a more careful understanding of how hadronic objects might be wrongly reconstructed or misidentified as lepton-jets. Muons on the other hand are reconstructed using the tracking chambers and this gives a better measurement of the kinematics. With this consideration we will also take the prompt lepton and lepton-jet to be muonic.

\begin{figure}
  \centering
   \includegraphics[width=0.5\textwidth]{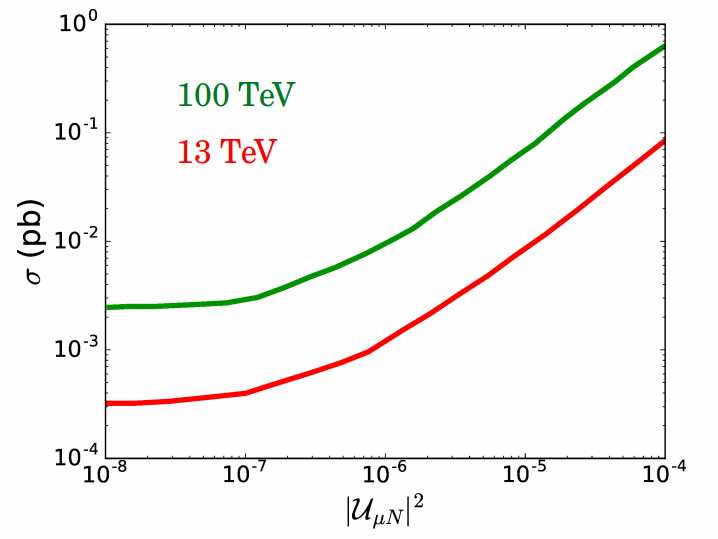}
  \caption{The production cross section for a sterile, right-handed neutrino in association with a lepton, generated via on-shell $W^\pm$ decay. The plot is for $m_{N_R}=8\GeV$ and the variation is
  shown as a function of the mixing. The two curves are for the $13\TeV$ (red) and $100\TeV$ (green) cases.}
  \label{lim8}
\end{figure}

In the $4 \GeV<m_{N_R} < 25\GeV$ mass-range, the dominant mode of production for $N_R$ is via an on-shell $W^\pm$ boson, as in \Fig{fig:signal_event}. The $N_R$ is produced in association with a prompt muon
\ba
W^{\pm}~\rightarrow~ N_R ~ + ~\mu^{\pm}\; .
\ea
The cross-sections for this production channel could differ by an order of magnitude between a $13\TeV$ and $100\TeV$ hadron collider. In \Fig{lim8} we illustrate this variation for the production cross-section in the case of $m_{N_R}=8\GeV$, as a function of the mixing. As we shall see, in the case of the backgrounds this increase can be even more drastic presenting challenges at $100\,\TeV$.

After production, we will consider the fully leptonic decay of the $N_R$, proceeding via an off-shell $W^{\pm *}$ or $Z^{0 *}$ boson 
\bea
N_R &\rightarrow& \mu^{\mp}+W^{\pm *} \rightarrow \mu^{\mp} + \mu^{\pm} + \nu_\mu \\ \nn
&\hookrightarrow& \nu_{\mu}~+~Z^{0 *} \rightarrow \nu^{\mu} + \mu^{\mp} + \mu^{\pm}\; .
\eea

We optimize our analysis for the case where the $N_R$ is boosted. This results in the final state muons and neutrino arising from its decay to be collimated. Thus our signal lepton-jet of specific interest is a muon-lepton-jet
\ba
\mu\text{-Jet} ~: ~ <\mu^{\pm}\mu^{\mp}\nu_\mu>\;
\ea
where the muons and neutrino are tightly collimated in a small cone-radius. We will require these pair of muons to be within a cone of radius $\Delta R<0.5$, where $\Delta R = \sqrt{\Delta\eta^2 + \Delta\phi^2}$. We will refer to this object as a muon-jet or $\mu\text{-Jet}$ henceforth.

Due to the boost of the $N_R$ and small mixing angles, the $\mu\text{-Jet}$ will be displaced from the prompt muon at the primary vertex. So, the characteristic signal being searched for consists of a prompt muon and a displaced $\mu\text{-Jet}$.

We generate the signal processes using \texttt{MadGraph5\_aMC@NLO}~\cite{Alwall:2014hca} for both $13\TeV$ and $100\TeV$. The parton showering and hadronization are preformed using \texttt{Pythia 8.219}\Cit{Sjostrand:2014zea} with \texttt{Tune 4C} used to simulate the busy hadronic environment. The hadronized output is then passed through the \texttt{Delphes 3.3.2}~\cite{deFavereau:2013fsa} detector simulation. We use the default CMS and FCC detector cards for $13\,\TeV$ and $100\,\TeV$ respectively. 

The dominant SM backgrounds arise from $W^\pm$ production in association with jets and from $\ttbar$ production. The $Z\rightarrow 4\ell$ and $WZ$ backgrounds are reduced
by demanding the invariant mass of the $\mu$-Jet with the prompt muon, $m_{\mu\text{-Jet}-\rm{prompt\,}\mu } < 80 \GeV$. The additional vetoing of a second prompt muon helps to completely
remove the contribution from $Z$, and $WZ$ processes. These selections do not impact the signal given the lack of a second prompt muon in the signal, and since the signal process begins with an
on-shell $W$ boson.

The background processes also follow the same simulation chain as the signal. The exact efficiencies of reconstructing non-standard objects, such as muon-jets, at a future $100\TeV$ detector are
of course less well understood, and must await a detailed description of the final detector design. We do our analysis using generator level hadronized output at both 13~\TeV\ and 100~\TeV.
To assess the effect of reconstruction efficiency on the signal, we consider two scenarios: a per-muon efficiency of 90\%, which will result in an event efficiency of about 70\%, and a per-muon
efficiency of 80\%, which will result in an event efficiency of about 50\%.
We start by making a selection for the prompt muon. We require the prompt muon to satisfy $\pt > 22\GeV$, and $\abseta < 2.4$. We then also make additional requirements on the impact parameter of the muon to ensure promptness, while requiring the muon to be isolated. At the LHC, this prompt isolated muon can be used to trigger the event. For the $\mu\text{-Jet}$, we start with selections based on the unique kinematics and topology of the signal, and subsequently impose further criteria that help to discriminate against dominant backgrounds, that may still contaminate the signal region. 

Overall, our signal selection criteria may be listed as:
\begin{itemize}
\item S0 : Require an isolated, prompt muon with $\pt> 22\GeV$, and $\abseta < 2.4$. Transverse impact parameter $d_{XY} < 0.2$mm and $d_{Z} < 0.1$mm.
The prompt muon is required to have the relative isolation, $\dfrac{\Sigma p_{\mathrm{trk}}}{\pt} < 25\%$. Here $\Sigma p_{\mathrm{trk}}$ is the sum of transverse momentum of all charged particles with $p_{\mathrm{trk}}> 1\GeV$ around a cone of $\Delta R < 0.4$ from the prompt muon.

\item S1: We require the $\mu\text{-Jet}$ to be composed of a pair of muons with opposite charge, and with $\pt>2~\GeV$ and $\abseta<2.4$. This pair of muons should also satisfy $\Delta R < 0.5$. The $\mu\text{-Jet}$ 4-vector is constructed by adding the 4-vectors of the two muons which form the $\mu\text{-Jet}$.
\item S2: We require the invariant mass of the $\mu\text{-Jet}$ with the prompt muon, $m_{\mu\text{-Jet}-\rm{prompt\,}\mu } < 80 \GeV$ . We also require that there is not more than one prompt muon per event. Both these requirements reduce the contribution from $Z\rightarrow 4\ell$ to negligible levels. In addition since the signal is produced starting with an on-shell $W$-boson, we also expect the invariant mass to not contribute beyond the $W$-boson mass.
\item S3: The signal does not have significant hadronic activity, while the primary backgrounds have jets. We require $H_{T}<60~\GeV$, where $H_T$ is defined as the scalar sum of $\pt$ of all AK4 jets in the event with
$\pt>30~\GeV$. This selection reduces both the $\ttbar$ and the $W^\pm$+Jets background.
\item S4: The azimuthal angle between the missing transverse energy (MET) and the $\mu\text{-Jet}$ should satisfy $\Delta \phi_{\rm{\text{muon-jet}}-\rm{MET}} < 0.5$. This selection supresses the $\ttbar$ background and the $W^\pm$+Jets background, where the $\Delta \phi_{\rm{\text{muon-jet}}-\rm{MET}}$ has no preferential value.
\item S5: We construct an isolation variable for the $\mu\text{-Jet}$ as the sum of transverse momenta of all charged tracks with $\pt>1\GeV$ within a cone of $\Delta R < 0.6$ from the $\mu\text{-Jet}$. We require this sum
to be less than $3\GeV$. This selection strongly discriminates against the $\ttbar$ and $W^\pm$+Jets background since the muons in these processes are accompanied by hadronic activity.
\end{itemize}

At $100\TeV$, the fraction of signal events that are produced in the forward direction increases as compared to $13\TeV$. The existing LHC experiments have coverage up to $\abseta<2.5$ for muons, and $\abseta<5$ for the calorimeters. We have considered that the detectors at a future $100\TeV$ collider will have extended muon coverage, as compared to present detectors, and thus we modify our selection to $\abseta<5.0$ for all muons in our $100\TeV$ analysis. But being in the narrow $4\gev <m_{N_R}<25\gev$ signal regime, we find that most kinematic quantities of interest, such as the $\pt$ of muons, MET etc. are quite similar between $13\TeV$ and $100\TeV$. We have therefore adopted, as evident from selection criteria S0-S5 earlier, identical selections for $13\TeV$  and $100\TeV$ studies. We have performed cross-checks to ensure the robustness of these assumptions.	
	
\begin{figure}
  \centering
   \includegraphics[width=0.5\textwidth]{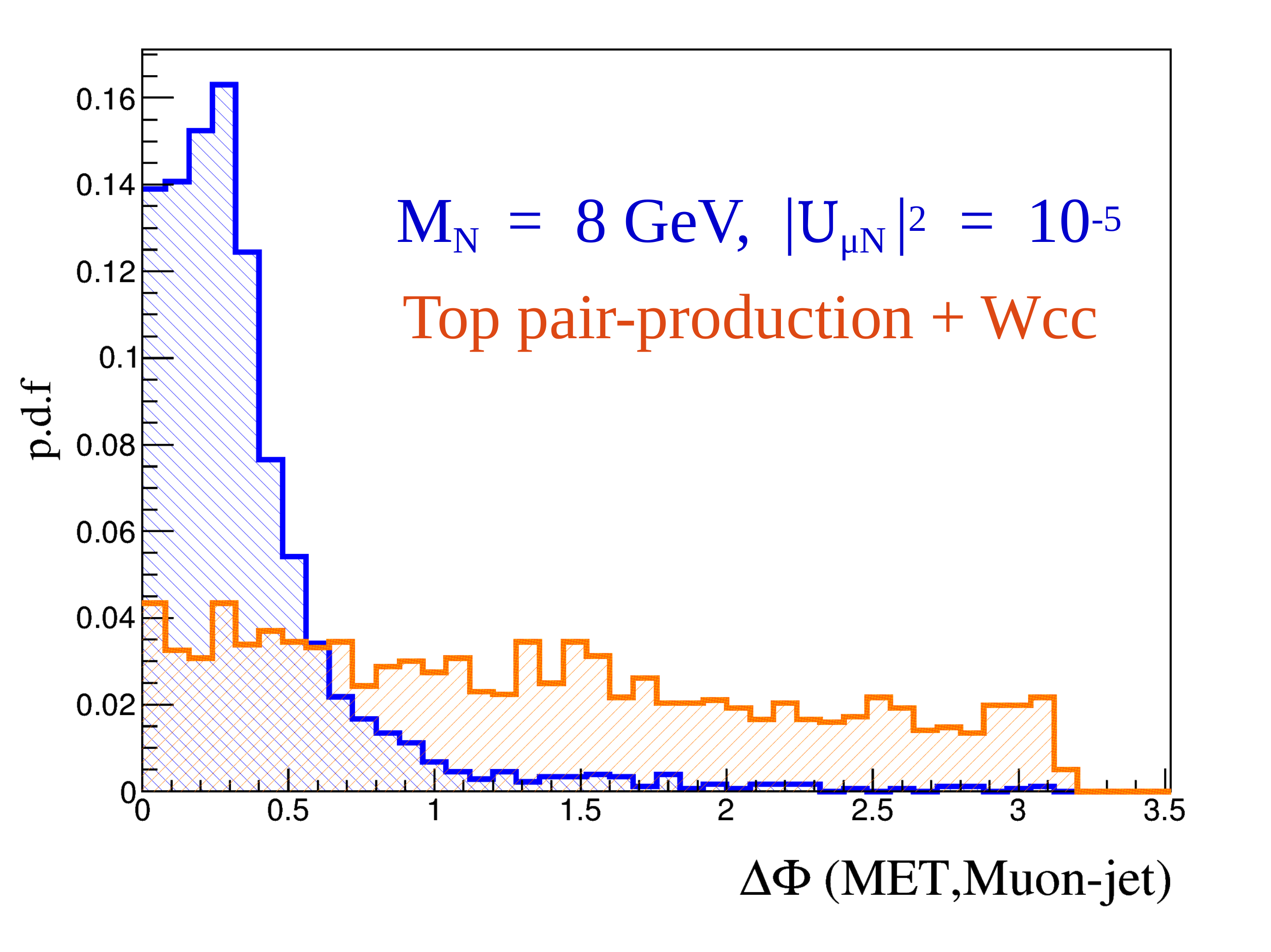}
   \includegraphics[width=0.5\textwidth]{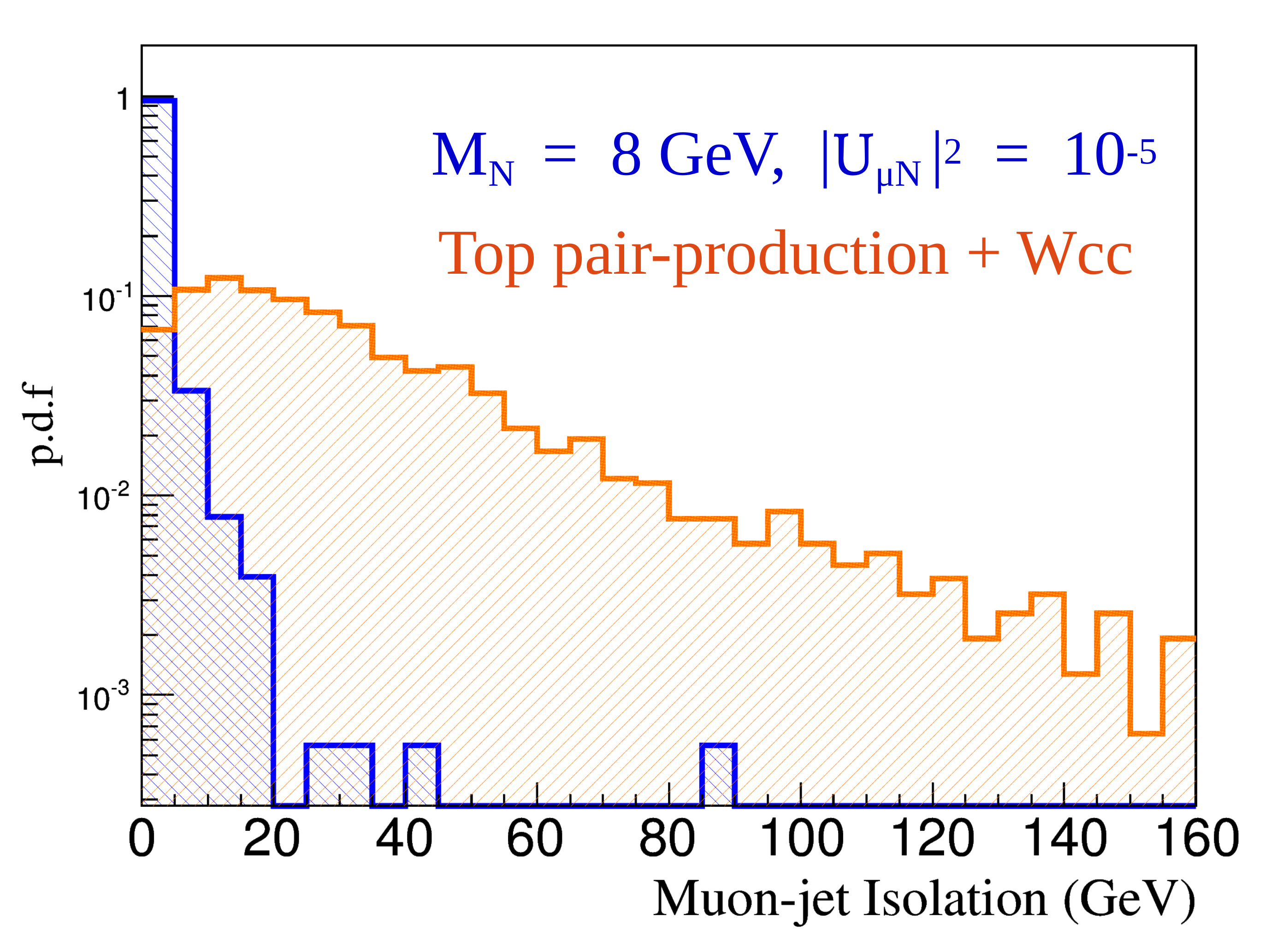}
  \caption{Depicted on top is the azimuthal angle between the missing transverse energy (MET) and the $\mu\text{-Jet}$ for signal and the combined background ($\ttbar$ and $W+$jets) for a 13~\TeV\ collider. As expected the $\Delta \phi_{\rm{\text{muon-jet}}-\rm{MET}}$ peaks at a lower value for the signal while showing no particular preference for any value for the backgrounds. Shown on the bottom is the isolation of the $\mu\text{-Jet}$ for the signal and combined background. The $\mu\text{-Jet}$ for the backgrounds is much lesser isolated due to busy hadronic activity. Both of these are obtained after the selections S0 through S2 have been imposed.}
  \label{13TeV_MET}
\end{figure}

In \Fig{13TeV_MET}(a) we show the azimuthal angle between the missing transverse energy (MET) and the $\mu\text{-Jet}$ for signal and the combined background ($\ttbar$ and $W+$jets) for a 13~\TeV\ collider. As expected the signal is concentrated at the lower end,
while the background has uniformly distributed value of $\Delta \phi_{\rm{\text{muon-jet}}-\rm{MET}}$ on average. \Fig{13TeV_MET}(b) shows the isolation of the $\mu\text{-Jet}$ of the signal and the combined background ($\ttbar$ and $W+$jets) for the 13 \TeV\ collider after our signal selections S0 through S2 have been imposed. The $\ttbar$ production ($t\rightarrow Wb\rightarrow \ell\nu b$) and the $W+$jets gives rise to a muons in a cascade decay. Hence due to their busy hadronic environment, the $\mu\text{-Jet}$ for the backgrounds are lesser isolated than the signal. We also consider cosmic-rays as a background. Given our topology a cosmic-ray can only act as a background if it passes through the interaction point (thus acting as the prompt muon in the event, and one of the $\mu\text{-Jet}$
muons). Following the estimate presented in  Ref.~\cite{Izaguirre:2015pga}, we consider this background to be negligible.

Previous studies have considered the impact-parameter and displacement of the $\mu\text{-Jet}$ muons to be a sharp discriminating variable, against background, and have made selections for displaced muon-jets. We find that placing too hard a cut on these variables actually reduces overall sensitivity to events where the $\mu\text{-Jet}$ is sometimes less displaced. It is also found that after other selection criteria it does not impact the remaining dominant $Wc\bar{c}$ or $t\bar{t}$ backgrounds significantly. The primary backgrounds arising from $\ttbar$ decay involve $b$-hadrons. These $b$-hadrons have lifetimes of order $c\tau\sim500\mu m$, resulting in muon displacement distributions that appear similar to signal over a significant part of the parameter space. Given these reasons, we do not actually make any hard impact-parameter requirements or displacement requirements for the $\mu\text{-Jet}$ muons.

Given the selections described above (S0 through S5), \Tab{sensEvt13} shows the acceptance for signal and background for our 13~\TeV\ analysis, while \Tab{sensEvt100} shows the same for the 100~\TeV\ analysis. It is evident that the veto on hadronic activity, and the $\Delta\phi$ requirement between the $\mu\text{-Jet}$ and MET reduces the background drastically while maintaining high signal sensitivity. As an alternative to the hadronic activity veto, as a cross-check, we also performed a separate study using $b$-tagging to assess the impact on the $\ttbar$ background. For this study, we considered the $b$-tagging efficiency from the CMS experiment \cite{CMS:2016btag}. We find, perhaps unsurprisingly, that an overall hadronic activity veto acts as a better background discriminant than using $b$-tags given the typical $b$-tagging efficiencies of 90\% with misidentification rate of about 1\%.

\begin{table}[h]
\caption{ The acceptance for signal and background for the 13~\TeV analysis. \label{sensEvt13} }
\begin{center}
\begin{tabular}{c  c  c  c }
  \hline \hline
   Selections & Signal & $\ttbar$ & Wcc  \\
  \hline
   S0:Acceptance [\%] & 47.8 & $22.6$ & $67.3$ \\
   S1:Acceptance [\%] & 18.9 & $3.5\times 10^{-1}$ & $2.1\times 10^{-2}$ \\
   S2:Acceptance [\%] & 17.9 & $2.6\times 10^{-1}$ & $1.6\times 10^{-2}$ \\
   S3:Acceptance [\%] & 16.6 & $6.7\times 10^{-4}$ & $10^{-2}$  \\
   S4:Acceptance [\%]  & 13.4 & $6.7\times 10^{-4}$ & $10^{-3}$  \\
   S5:Acceptance [\%]  & 12.2 & $5\times 10^{-6}$ & $2.3\times 10^{-4}$  \\  
  \hline
\hline
\end{tabular}
\end{center}
\end{table}       

\begin{table}[h]
\caption{ The acceptance for signal and background for the 100~\TeV analysis. \label{sensEvt100}}
\begin{center}
\begin{tabular}{c  c  c  c }
  \hline \hline
   Selections & Signal & $\ttbar$ & Wcc  \\
  \hline
   S0:Acceptance [\%] & 64.3 & $30.6$ & $85.1$ \\
   S1:Acceptance [\%] & 16.0 & $4.1\times 10^{-1}$ & $3.2\times 10^{-2}$ \\
   S2:Acceptance [\%] & 11.9 & $2.5\times 10^{-1}$ & $2.5\times 10^{-2}$ \\
   S3:Acceptance [\%] & 9.2 & $7.4\times 10^{-3}$ & $6\times 10^{-3}$  \\
   S4:Acceptance [\%]  & 7.2 & $3\times 10^{-4}$ & $2\times 10^{-3}$  \\
   S5:Acceptance [\%]  & 6.8 & $3.5\times 10^{-5}$ & $3.1\times 10^{-4}$  \\  
  \hline
\hline
\end{tabular}
\end{center}
\end{table}

In \Fig{excl13n100}, we compare the final estimated sensitivity for our selections. Existing constraints are shown as dotted curves. The contours are for 13 \TeV\ LHC with 300 \ifb\ data (red)
and 100 \TeV\ FCC-hh/SppC also assuming 300 \ifb\ data (green). We show 95\% CL limits on \mix as a function of right-handed neutrino masses $m_{N_R}$. The limits were computed using the
asymptotic limit method \cite{Junk:1999kv}-\cite{CMS-NOTE-2011-005}.  Assuming a 100\% event reconstruction
efficiency for signal, the upper limit on the signal cross-section is calculated to be $\sigma_{\text{\tiny{LIM}}}= 9.03\times 10^{-4} \,\rm{pb}$. If we consider efficiencies of 70\% and 50\%, the
upper limit worsens to $1.3\times 10^{-3}$ pb and $1.8\times 10^{-3}$ pb respectively. As expected, the reach of the experiment will depend on the efficiency with which muons (and the signal events) are reconstructed. Note that at
higher $m_{N_R}$, the sensitivity decreases since the daughter muons no longer satisfy the geometric criteria for a $\mu\text{-Jet}$. A high sensitivity at low $m_{N_R}$ is maintained due to the extreme low momenta muons considered here.

\begin{figure}
  \centering
   \includegraphics[width=0.485\textwidth]{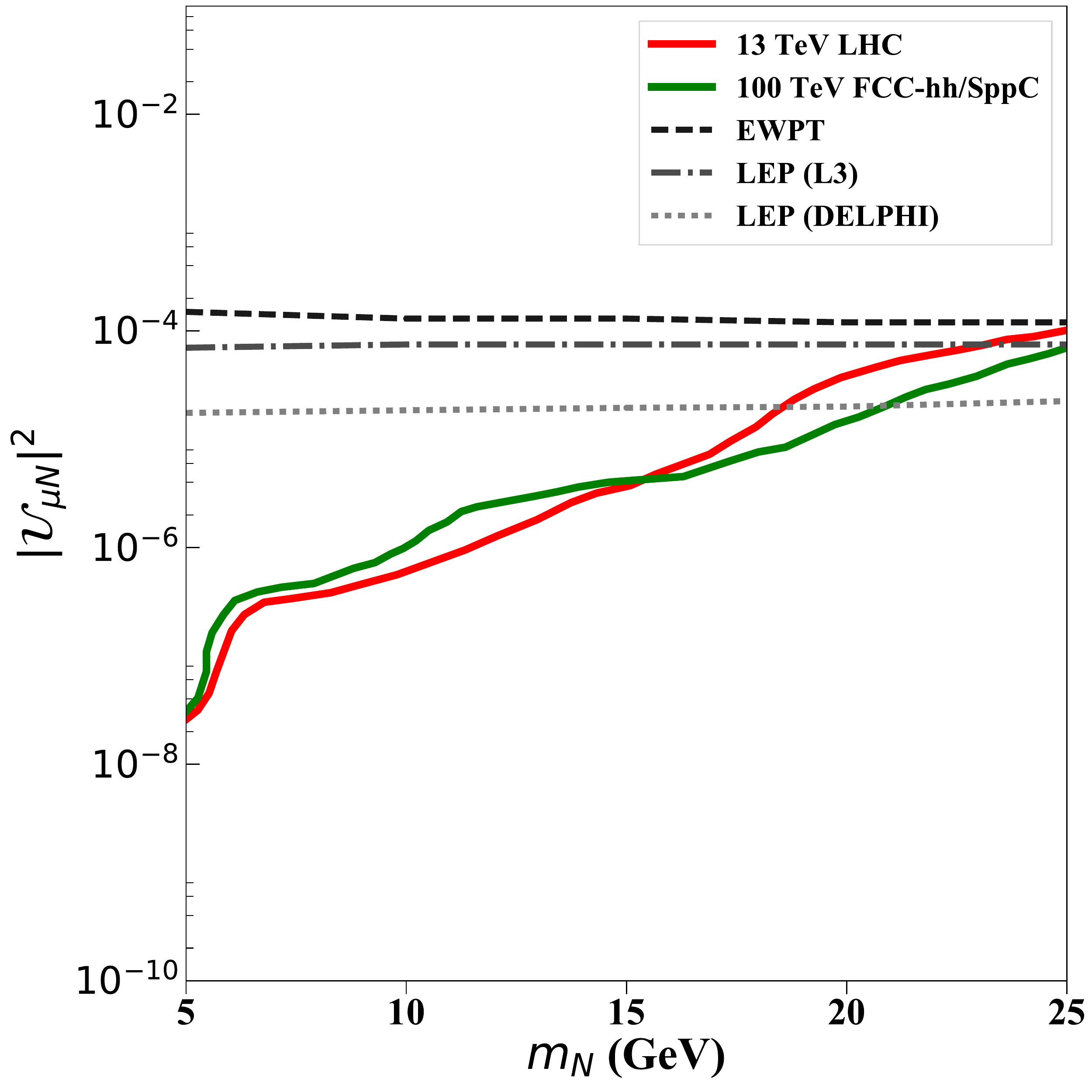}
  \caption{Exclusion contours for the $13 \TeV$ LHC (red) and the proposed $100\TeV$ FCC-hh/SppC (green), for similar selection criteria, assuming a signal efficiency of $100\%$. For the same integrated luminosity of $300\rm{fb}^{-1}$, and for a similar cut based analysis, the LHC is already competent with a $100\TeV$ collider. Constraints from electroweak precision data (EWPT)\Cit{delAguila:2008pw, Akhmedov:2013hec, Basso:2013jka, Blas:2013ana, Antusch:2015mia} and LEP data (L3 and DELPHI collaborations)\Cit{Adriani:1992pq,Abreu:1996pa} are also shown for comparisons.}
  \label{excl13n100}  
\end{figure}

As shown in \Fig{excl13n100}, the $13\TeV$ LHC search optimised for this mass-regime is already competitive in sensitivity, if not slightly better in some regions, to the sensitivity of the proposed $100\TeV$ hadron collider. This may seem surprising and contrary to naive expectations. The primary reason is again the narrow $[4\,\rm{GeV},\,25\,\rm{GeV}]$ signal region that we are trying to optimize over and the presence of strongly produced backgrounds that disproportionately increase when one moves from $13\TeV$ to $100\TeV$. Even though the signal cross-section of our signal increases by an order of magnitude at the $100\TeV$ collider, as evidenced by Fig.\,\ref{lim8}, we observe that the background cross-section increases even more drastically -- about 5 times as rapidly as the signal. This is of course expected as backgrounds like top-pair production feed off the strong production modes while our signal is dominantly produced solely via the weak interaction, in this regime. The drastic increase in background cross-sections at $100\TeV$ renders some of the signal cross-section increase and selection optimisations impotent. Thus, if our analysis and understanding is correct, for a $100\TeV$ hadron collider to do significantly better than the LHC, an increased detector coverage and algorithmic improvements, such as in b-tagging, might be required, along with more sophisticated search strategies. Based on our study, our current conclusion is therefore that the $13\TeV$ LHC, for the final states of interest, can give competitive limits in the $[4\,\rm{GeV},\,25\,\rm{GeV}]$ sterile neutrino mass-regime.

%------------------------------------------------------------------------------------------------------------------
\section {\label{sec:5} Summary and Conclusions}
\label{sec:5}
Sterile neutrino states are well motivated in many extensions of the Standard Model and have the potential to cast light on many unsolved questions in our theoretical frameworks. With the realization that their masses are not \textit{a priori} fixed to any particular mass-scale, it becomes crucial to have search strategies spanning all the possible values. 

We focused on a relatively unexplored mass regime ($\Lambda_{\text{\tiny{QCD}}} \ll m_{N_R} \ll m_{W^\pm}$), where current constraints and experimental searches at hadron colliders are lacking. Also, prior theoretical studies in this signal region seem to have missed certain subtle, albeit crucial aspects of backgrounds and selection, while making sensitivity projections. 

Motivated by the previous studies, unique signal topology and challenges singular to hadron colliders, we specifically revisited the sterile neutrino signal topology consisting of a prompt lepton and a displaced lepton-jet. We have attempted to make the first systematic study in this signal region, for the $13\,\rm{TeV}$ high-luminosity Large Hadron Collider and a future FCC-hh/SppC $100\,\rm{TeV}$ $pp$-collider. 

For the same set of selection cuts, albeit for selection criteria optimised to each collider, our conclusion is that the $13\,\rm{TeV}$ HL- LHC may already be competitive with a future hadron collider. This is partially due to the fact that we are optimizing over a narrow mass-region in the low, electroweak regime for the signal -- so the gains in signal cross-section while going to a higher energy machine are moderate -- while the relevant backgrounds for the toplogy under consideration increase much more drastically.  A higher detector coverage, algorithmic improvements and a more sophisticated search strategy, in contrast to the simple cut and count based analysis we have performed, may possibly improve the reach at a $100\,\rm{TeV}$ $pp$-collider significantly. On the other hand a future $e^+e^-$ collider may be able to significantly extend the sensitivities to very low mixing angles (see for instance\Cit{Antusch:2016ejd} and references therein). Also, during the completion of this work an interesting study\Cit{Antusch:2017hhu} appeared that looks for sterile neutrinos by reinterpreting displaced vertex searches ($\mu j j$ final states) for long-lived particles at LHCb, during run-1\Cit{Aaij:2016xmb}. The study recasts current data and makes projections for future LHCb searches and the final limits are comparable to our study in the mass regime of interest; nevertheless with a different exclusion-limit functional profile.

A systematic and continuing program of sterile neutrino searches at current and future colliders, in all relevant mass ranges and topologies, would help in elucidating the nature of these states, if they exist, and help towards a complete coverage of interesting signal regions.

%------------------------------------------------------------------------------------------------------------------
\section*{Acknowledgements}
We would like to thank S. Chauhan, E. Izaguirre, B. Shuve, S. Somalwar and S. Thomas for discussions and comments on the manuscript. S.D. and D.G are grateful to E. Izaguirre and B. Shuve for providing model files from their earlier study for corroboration and to S. Chauhan for help with simulation. The computations for the study were performed at the High Power Computing facility of the EHEP group at IISER, Pune.

%------------------------------------------------------------------------------------------------------------------
\bibliography{RHN_LeptonJets}

\providecommand{\href}[2]{#2}\begingroup\raggedright\begin{thebibliography}{10}


%%%%%%%%%%%%%%%%%%% LHC & Higgs %%%%%%%%%%%%%%%%%%%%%%%%
 %\cite{Chatrchyan:2012ufa}
\bibitem{Chatrchyan:2012ufa} 
  S.~Chatrchyan {\it et al.}  [CMS Collaboration],
  ``Observation of a new boson at a mass of 125 GeV with the CMS experiment at the LHC,''
  Phys.\ Lett.\ B {\bf 716}, 30 (2012)
  [arXiv:1207.7235 [hep-ex]].
  %%CITATION = ARXIV:1207.7235;%%

%\cite{Aad:2012tfa}
\bibitem{Aad:2012tfa} 
  G.~Aad {\it et al.}  [ATLAS Collaboration],
  ``Observation of a new particle in the search for the Standard Model Higgs boson 
  with the ATLAS detector at the LHC,''
  Phys.\ Lett.\ B {\bf 716}, 1 (2012)
  [arXiv:1207.7214 [hep-ex]].
  %%CITATION = ARXIV:1207.7214;%%
  
  
  %%%%%%%%%%%% Family symmetries - Mass & Mixing%%%%%%%%%%%%%%
  
  %\cite{McKeen:2007ry}
\bibitem{McKeen:2007ry}
  D.~McKeen, J.~L.~Rosner and A.~M.~Thalapillil,
  %``Masses and Mixings in a Grand Unified Toy Model,''
  Phys.\ Rev.\ D {\bf 76} (2007) 073014
  doi:10.1103/PhysRevD.76.073014
  [hep-ph/0703177].
  %%CITATION = doi:10.1103/PhysRevD.76.073014;%%
  %6 citations counted in INSPIRE as of 09 May 2017
  
  %%%%%%%%%% Sterile neutrino indications in oscillation expts. %%%%%%%%%%%%
  
  %\cite{Aguilar:2001ty}
\bibitem{Aguilar:2001ty}
  A.~Aguilar {\it et al.} [ LSND Collaboration ],
  %``Evidence for neutrino oscillations from the observation of anti-neutrino(electron) appearance in a anti-neutrino(muon) beam,''
  Phys.\ Rev.\  {\bf D64}, 112007 (2001).
  [hep-ex/0104049].
  
  
   %\cite{AguilarArevalo:2010wv}
\bibitem{AguilarArevalo:2010wv}
  A.~A.~Aguilar-Arevalo {\it et al.} [ The MiniBooNE Collaboration ],
  %``Event Excess in the MiniBooNE Search for $\bar \nu_\mu \rightarrow \bar \nu_e$ Oscillations,''
  Phys.\ Rev.\ Lett.\  {\bf 105}, 181801 (2010).
  [arXiv:1007.1150 [hep-ex]]
  
 \bibitem{MiniBooNE_new_antineu}  
 E. Zimmerman [ MiniBooNE Collaboration ], PANIC 2011; Z. Djurcic [ MiniBooNE Collaboration ] NUFACT 2011; E.~D.~Zimmerman [ MiniBooNE Collaboration ],
  %``Updated Search for Electron Antineutrino Appearance at MiniBooNE,''
  [arXiv:1111.1375 [hep-ex]].
  
  
  %\cite{Mention:2011rk}
\bibitem{Mention:2011rk}
  G.~Mention, M.~Fechner, T.~.Lasserre, T.~.A.~Mueller, D.~Lhuillier, M.~Cribier, A.~Letourneau,
  %``The Reactor Antineutrino Anomaly,''
  Phys.\ Rev.\  {\bf D83}, 073006 (2011).
  [arXiv:1101.2755 [hep-ex]].  
  
   
  %%%%%%%%% Sterile Neutrino considerations in oscillations %%%%%%%%%%%%%%%%
  
  
  %\cite{Donini:2001xy}
\bibitem{Donini:2001xy} 
  A.~Donini and D.~Meloni,
  %``The 2+2 and 3+1 four family neutrino mixing at the neutrino factory,''
  Eur.\ Phys.\ J.\ C {\bf 22}, 179 (2001)
  [hep-ph/0105089].
  %%CITATION = HEP-PH/0105089;%%
  

  %\cite{Sorel:2003hf}
\bibitem{Sorel:2003hf}
  M.~Sorel, J.~M.~Conrad, M.~Shaevitz,
  %``A Combined analysis of short baseline neutrino experiments in the (3+1) and (3+2) sterile neutrino oscillation hypotheses,''
  Phys.\ Rev.\  {\bf D70}, 073004 (2004).
  [hep-ph/0305255].
     
   
  %\cite{Dighe:2007uf}
\bibitem{Dighe:2007uf} 
  A.~Dighe and S.~Ray,
  %``Signatures of heavy sterile neutrinos at long baseline experiments,''
  Phys.\ Rev.\ D {\bf 76}, 113001 (2007)
  [arXiv:0709.0383 [hep-ph]].
  %%CITATION = ARXIV:0709.0383;%%
  
  %\cite{Donini:2007yf}
\bibitem{Donini:2007yf} 
  A.~Donini, M.~Maltoni, D.~Meloni, P.~Migliozzi and F.~Terranova,
  %``3+1 sterile neutrinos at the CNGS,''
  JHEP {\bf 0712}, 013 (2007)
  [arXiv:0704.0388 [hep-ph]].
  %%CITATION = ARXIV:0704.0388;%%
  
  %\cite{Maltoni:2007zf}
\bibitem{Maltoni:2007zf} 
  M.~Maltoni and T.~Schwetz,
  %``Sterile neutrino oscillations after first MiniBooNE results,''
  Phys.\ Rev.\ D {\bf 76}, 093005 (2007)
  doi:10.1103/PhysRevD.76.093005
  [arXiv:0705.0107 [hep-ph]].
  %%CITATION = doi:10.1103/PhysRevD.76.093005;%%
  %261 citations counted in INSPIRE as of 18 May 2017
  
 
 %\cite{Donini:2008wz}
\bibitem{Donini:2008wz} 
  A.~Donini, K.~-i.~Fuki, J.~Lopez-Pavon, D.~Meloni and O.~Yasuda,
  %``The Discovery channel at the Neutrino Factory: nu(mu) ---> nu(tau) pointing to sterile neutrinos,''
  JHEP {\bf 0908}, 041 (2009)
  [arXiv:0812.3703 [hep-ph]].
  %%CITATION = ARXIV:0812.3703;%%
  
  %\cite{deGouvea:2008qk}
\bibitem{deGouvea:2008qk} 
  A.~de Gouvea and T.~Wytock,
  %``Light Sterile Neutrino Effects at theta(3)-Sensitive Reactor Neutrino Experiments,''
  Phys.\ Rev.\ D {\bf 79}, 073005 (2009)
  doi:10.1103/PhysRevD.79.073005
  [arXiv:0809.5076 [hep-ph]].
  %%CITATION = doi:10.1103/PhysRevD.79.073005;%%
  %30 citations counted in INSPIRE as of 18 May 2017
    
  
  %\cite{Meloni:2010zr}
\bibitem{Meloni:2010zr} 
  D.~Meloni, J.~Tang and W.~Winter,
  %``Sterile neutrinos beyond LSND at the Neutrino Factory,''
  Phys.\ Rev.\ D {\bf 82}, 093008 (2010)
  [arXiv:1007.2419 [hep-ph]].
  %%CITATION = ARXIV:1007.2419;%%
  
  %\cite{Nelson:2010hz}
\bibitem{Nelson:2010hz}
  A.~E.~Nelson,
  %``Effects of CP Violation from Neutral Heavy Fermions on Neutrino
  %Oscillations, and the LSND/MiniBooNE Anomalies,''
  Phys.\ Rev.\  D {\bf 84}, 053001 (2011)
  [arXiv:1010.3970 [hep-ph]].
  %%CITATION = PHRVA,D84,053001;%%
  
  %\cite{Kopp:2011qd}
\bibitem{Kopp:2011qd} 
  J.~Kopp, M.~Maltoni and T.~Schwetz,
  %``Are there sterile neutrinos at the eV scale?,''
  Phys.\ Rev.\ Lett.\  {\bf 107}, 091801 (2011)
  doi:10.1103/PhysRevLett.107.091801
  [arXiv:1103.4570 [hep-ph]].
  %%CITATION = doi:10.1103/PhysRevLett.107.091801;%%
  %253 citations counted in INSPIRE as of 18 May 2017
  
  %\cite{Karagiorgi:2011ut}
\bibitem{Karagiorgi:2011ut} 
  G.~Karagiorgi,
  %``Confronting Recent Neutrino Oscillation Data with Sterile Neutrinos,''
  arXiv:1110.3735 [hep-ph].
  %%CITATION = ARXIV:1110.3735;%%
  %18 citations counted in INSPIRE as of 18 May 2017
  
  %\cite{Bhattacharya:2011ee}
\bibitem{Bhattacharya:2011ee} 
  B.~Bhattacharya, A.~M.~Thalapillil and C.~E.~M.~Wagner,
  %``Implications of sterile neutrinos for medium/long-baseline neutrino experiments and the determination of $\theta_{13}$,''
  Phys.\ Rev.\ D {\bf 85}, 073004 (2012)
  doi:10.1103/PhysRevD.85.073004
  [arXiv:1111.4225 [hep-ph]].
  %%CITATION = doi:10.1103/PhysRevD.85.073004;%%
  %32 citations counted in INSPIRE as of 18 May 2017
  
  %\cite{Giunti:2011gz}
\bibitem{Giunti:2011gz} 
  C.~Giunti and M.~Laveder,
  %``3+1 and 3+2 Sterile Neutrino Fits,''
  Phys.\ Rev.\ D {\bf 84}, 073008 (2011)
  doi:10.1103/PhysRevD.84.073008
  [arXiv:1107.1452 [hep-ph]].
  %%CITATION = doi:10.1103/PhysRevD.84.073008;%%
  %156 citations counted in INSPIRE as of 18 May 2017
  
  %\cite{Giunti:2011ht}
\bibitem{Giunti:2011ht}
  C.~Giunti,
  %``Sterile Neutrino Fits,''
  [arXiv:1106.4479 [hep-ph]]; C.~Giunti, M.~Laveder,
  %``3+1 and 3+2 Sterile Neutrino Fits,''
  [arXiv:1107.1452 [hep-ph]].
  
  %\cite{Barger:2011rc}
\bibitem{Barger:2011rc}
  V.~Barger, Y.~Gao, D.~Marfatia,
  %``Is there evidence for sterile neutrinos in IceCube data?,''
   [arXiv:1109.5748 [hep-ph]].
   
  %%%%%%%%%%%%%%%%%%%%%%%%% Daya bay reactor explanation %%%%%%%%%%%%%%%%%%%%%%%%%
   
    %\cite{An:2017osx}
\bibitem{An:2017osx}
  F.~P.~An {\it et al.} [Daya Bay Collaboration],
  %``Evolution of the Reactor Antineutrino Flux and Spectrum at Daya Bay,''
  %Submitted to: Phys.Rev.Lett.
  [arXiv:1704.01082 [hep-ex]].
  %%CITATION = ARXIV:1704.01082;%%
  %3 citations counted in INSPIRE as of 16 May 2017
   
%%%%%%%%%%%%%%%% Neutrino mass models %%%%%%%%%%%%%%%%%%%%%%%%%

%\cite{see-saw}
  \bibitem{see-saw}{P. Minkowski, Phys. Lett. B \textbf{67}, 421 (1977); M. Gell-Mann, P. Ramond and R. Slansky, \textit{ Supergravity}, North Holland, Amsterdam, 315 (1980); T. Yanagida, in \textit{Proceedings of the Workshop on the Unified Theory and the Baryon Number in the Universe}, KEK, Tsukuba, Japan, 95 (1979); S. L. Glashow, in \textit{Proceedings of the 1979 Cargese Summer Institute on Quarks and Leptons}, Plenum Press, New York, 687 (1980); R. N. Mohapatra and G. Senjanovic, Phys. Rev. Lett. \textbf{44}, 912 (1980); J. Schechter and J. W. F. Valle, Phys. Rev. D {\bf 22}, 2227 (1980); J. Schechter and J. W. F. Valle, Phys. Rev. D {\bf 25}, 774 (1982).}
  
\bibitem{Mohapatra:1974hk}
R.~N. Mohapatra and J.~C. Pati,
\newblock Phys.Rev. {\bf D11}, 566 (1975).

\bibitem{Mohapatra:1974gc}
R.~N. Mohapatra and J.~C. Pati,
\newblock Phys.Rev. {\bf D11}, 2558 (1975).

\bibitem{Senjanovic:1975rk}
G.~Senjanovic and R.~N. Mohapatra,
\newblock Phys.Rev. {\bf D12}, 1502 (1975).

%\cite{Witten:1979nr}
\bibitem{Witten:1979nr}
  E.~Witten,
  %``Neutrino Masses in the Minimal O(10) Theory,''
  Phys.\ Lett.\  {\bf 91B} (1980) 81.
  doi:10.1016/0370-2693(80)90666-8
  %%CITATION = doi:10.1016/0370-2693(80)90666-8;%%
  %371 citations counted in INSPIRE as of 16 May 2017

\bibitem{Nandi:1985uh}
S.~Nandi and U.~Sarkar,
\newblock Phys.Rev.Lett. {\bf 56}, 564 (1986).

\bibitem{mohapatra:1986bd}
R.~N. Mohapatra and J.~W.~F. Valle,
\newblock Phys. Rev. {\bf D34}, 1642 (1986).

\bibitem{mohapatra:1986aw}
R.~N. Mohapatra,
\newblock Phys. Rev. Lett. {\bf 56}, 561 (1986).


\bibitem{Babu:1992ia}
K.~S. Babu and R.~N. Mohapatra,
\newblock Phys.Rev.Lett. {\bf 70}, 2845 (1993), hep-ph/9209215.


\bibitem{Bajc:2006ia}
B.~Bajc and G.~Senjanovic,
\newblock JHEP {\bf 08}, 014 (2007), hep-ph/0612029.

  
  %\cite{Asaka:2005an}
\bibitem{Asaka:2005an}
  T.~Asaka, S.~Blanchet and M.~Shaposhnikov,
  %``The nuMSM, dark matter and neutrino masses,''
  Phys.\ Lett.\ B {\bf 631} (2005) 151
  doi:10.1016/j.physletb.2005.09.070
  [hep-ph/0503065].
  %%CITATION = doi:10.1016/j.physletb.2005.09.070;%%
  %459 citations counted in INSPIRE as of 16 May 2017
  
  %\cite{Asaka:2005pn}
\bibitem{Asaka:2005pn}
  T.~Asaka and M.~Shaposhnikov,
  %``The nuMSM, dark matter and baryon asymmetry of the universe,''
  Phys.\ Lett.\ B {\bf 620} (2005) 17
  doi:10.1016/j.physletb.2005.06.020
  [hep-ph/0505013].
  %%CITATION = doi:10.1016/j.physletb.2005.06.020;%%
  %433 citations counted in INSPIRE as of 16 May 2017
  
  %\cite{Canetti:2012vf}
\bibitem{Canetti:2012vf}
  L.~Canetti, M.~Drewes and M.~Shaposhnikov,
  %``Sterile Neutrinos as the Origin of Dark and Baryonic Matter,''
  Phys.\ Rev.\ Lett.\  {\bf 110} (2013) no.6,  061801
  doi:10.1103/PhysRevLett.110.061801
  [arXiv:1204.3902 [hep-ph]].
  %%CITATION = doi:10.1103/PhysRevLett.110.061801;%%
  %123 citations counted in INSPIRE as of 16 May 2017
  
  
  %\cite{Canetti:2012kh}
\bibitem{Canetti:2012kh}
  L.~Canetti, M.~Drewes, T.~Frossard and M.~Shaposhnikov,
  %``Dark Matter, Baryogenesis and Neutrino Oscillations from Right Handed Neutrinos,''
  Phys.\ Rev.\ D {\bf 87} (2013) 093006
  doi:10.1103/PhysRevD.87.093006
  [arXiv:1208.4607 [hep-ph]].
  %%CITATION = doi:10.1103/PhysRevD.87.093006;%%
  %173 citations counted in INSPIRE as of 16 May 2017
  
  %%%%%%%%%%%%%%%%%%%%%%%%%%%%%%%%%%%%%%%%
  
  %%%%%%%%%%%%%%% RH-Neutrino mass-scale and technical naturalness %%%%%%%%%%%%%

%\cite{Fujikawa:2004jy}
\bibitem{Fujikawa:2004jy}
  K.~Fujikawa,
  %``Remark on natural models of neutrinos,''
  Prog.\ Theor.\ Phys.\  {\bf 113} (2005) 1065
  doi:10.1143/PTP.113.1065
  [hep-ph/0407331].
  %%CITATION = doi:10.1143/PTP.113.1065;%%
  %9 citations counted in INSPIRE as of 11 May 2017
  
  
   %\cite{deGouvea:2005er}
\bibitem{deGouvea:2005er}
  A.~de Gouvea,
  %``See-saw energy scale and the LSND anomaly,''
  Phys.\ Rev.\ D {\bf 72} (2005) 033005
  doi:10.1103/PhysRevD.72.033005
  [hep-ph/0501039].
  %%CITATION = doi:10.1103/PhysRevD.72.033005;%%
  %106 citations counted in INSPIRE as of 11 May 2017
  
  %%%%%%%%% References added as advised by referee %%%%%%%%%%%
  
  %\cite{Keung:1983uu}
\bibitem{Keung:1983uu} 
  W.~Y.~Keung and G.~Senjanovic,
  %``Majorana Neutrinos and the Production of the Right-handed Charged Gauge Boson,''
  Phys.\ Rev.\ Lett.\  {\bf 50}, 1427 (1983).
  doi:10.1103/PhysRevLett.50.1427
  %%CITATION = doi:10.1103/PhysRevLett.50.1427;%%
  %365 citations counted in INSPIRE as of 29 Aug 2017
  
  %\cite{Pilaftsis:1991ug}
\bibitem{Pilaftsis:1991ug} 
  A.~Pilaftsis,
  %``Radiatively induced neutrino masses and large Higgs neutrino couplings in the standard model with Majorana fields,''
  Z.\ Phys.\ C {\bf 55}, 275 (1992)
  doi:10.1007/BF01482590
  [hep-ph/9901206].
  %%CITATION = doi:10.1007/BF01482590;%%
  %266 citations counted in INSPIRE as of 29 Aug 2017
  
  %\cite{Datta:1993nm}
\bibitem{Datta:1993nm} 
  A.~Datta, M.~Guchait and A.~Pilaftsis,
  %``Probing lepton number violation via majorana neutrinos at hadron supercolliders,''
  Phys.\ Rev.\ D {\bf 50}, 3195 (1994)
  doi:10.1103/PhysRevD.50.3195
  [hep-ph/9311257].
  %%CITATION = doi:10.1103/PhysRevD.50.3195;%%
  %144 citations counted in INSPIRE as of 29 Aug 2017


  %%%%%%%%%% Neutrino Collider & Non-Collider Reviews %%%%%%%%%%
  
  %\cite{delAguila:2006bda}
\bibitem{delAguila:2006bda} 
  F.~del Aguila, J.~A.~Aguilar-Saavedra and R.~Pittau,
  %``Neutrino physics at large colliders,''
  J.\ Phys.\ Conf.\ Ser.\  {\bf 53}, 506 (2006)
  doi:10.1088/1742-6596/53/1/032
  [hep-ph/0606198].
  %%CITATION = doi:10.1088/1742-6596/53/1/032;%%
  %50 citations counted in INSPIRE as of 18 May 2017
  
   %\cite{Atre:2009rg}
\bibitem{Atre:2009rg}
  A.~Atre, T.~Han, S.~Pascoli and B.~Zhang,
  %``The Search for Heavy Majorana Neutrinos,''
  JHEP {\bf 0905} (2009) 030
  doi:10.1088/1126-6708/2009/05/030
  [arXiv:0901.3589 [hep-ph]].
  %%CITATION = doi:10.1088/1126-6708/2009/05/030;%%
  %374 citations counted in INSPIRE as of 14 May 2017
  
    
  %\cite{Deppisch:2015qwa}
\bibitem{Deppisch:2015qwa} 
  F.~F.~Deppisch, P.~S.~Bhupal Dev and A.~Pilaftsis,
  %``Neutrinos and Collider Physics,''
  New J.\ Phys.\  {\bf 17}, no. 7, 075019 (2015)
  doi:10.1088/1367-2630/17/7/075019
  [arXiv:1502.06541 [hep-ph]].
  %%CITATION = doi:10.1088/1367-2630/17/7/075019;%%
  %141 citations counted in INSPIRE as of 18 May 2017
  
  
  %\cite{Dev:2016gvv}
\bibitem{Dev:2016gvv} 
  P.~S.~B.~Dev,
  %``Testing Neutrino Mass Models at the LHC and beyond,''
  PoS ICHEP {\bf 2016}, 487 (2016)
  [arXiv:1612.08209 [hep-ph]].
  %%CITATION = ARXIV:1612.08209;%%
  %2 citations counted in INSPIRE as of 28 May 2017
  
  %\cite{Antusch:2016ejd}
\bibitem{Antusch:2016ejd} 
  S.~Antusch, E.~Cazzato and O.~Fischer,
  %``Sterile neutrino searches at future $e^-e^+$, $pp$, and $e^-p$ colliders,''
  arXiv:1612.02728 [hep-ph].
  %%CITATION = ARXIV:1612.02728;%%
  %9 citations counted in INSPIRE as of 18 May 2017
  
  %%%%%%%%%%%%%%%% References added %%%%%%%%%%%%%%%%
  %\cite{Lindner:2016lxq}
\bibitem{Lindner:2016lxq} 
  M.~Lindner, F.~S.~Queiroz, W.~Rodejohann and C.~E.~Yaguna,
  %``Left-Right Symmetry and Lepton Number Violation at the Large Hadron Electron Collider,''
  JHEP {\bf 1606}, 140 (2016)
  doi:10.1007/JHEP06(2016)140
  [arXiv:1604.08596 [hep-ph]].
  %%CITATION = doi:10.1007/JHEP06(2016)140;%%
  %24 citations counted in INSPIRE as of 29 Aug 2017
  
  %\cite{Campos:2017odj}
\bibitem{Campos:2017odj} 
  M.~D.~Campos, F.~S.~Queiroz, C.~E.~Yaguna and C.~Weniger,
  %``Search for right-handed neutrinos from dark matter annihilation with gamma-rays,''
  JCAP {\bf 1707}, no. 07, 016 (2017)
  doi:10.1088/1475-7516/2017/07/016
  [arXiv:1702.06145 [hep-ph]].
  %%CITATION = doi:10.1088/1475-7516/2017/07/016;%%
  %7 citations counted in INSPIRE as of 29 Aug 2017
  
  %\cite{Abada:2012mc}
\bibitem{Abada:2012mc} 
  A.~Abada, D.~Das, A.~M.~Teixeira, A.~Vicente and C.~Weiland,
  %``Tree-level lepton universality violation in the presence of sterile neutrinos: impact for $R_K$ and $R_\pi$,''
  JHEP {\bf 1302}, 048 (2013)
  doi:10.1007/JHEP02(2013)048
  [arXiv:1211.3052 [hep-ph]].
  %%CITATION = doi:10.1007/JHEP02(2013)048;%%
  %71 citations counted in INSPIRE as of 29 Aug 2017
  
  %\cite{Abada:2013aba}
\bibitem{Abada:2013aba} 
  A.~Abada, A.~M.~Teixeira, A.~Vicente and C.~Weiland,
  %``Sterile neutrinos in leptonic and semileptonic decays,''
  JHEP {\bf 1402}, 091 (2014)
  doi:10.1007/JHEP02(2014)091
  [arXiv:1311.2830 [hep-ph]].
  %%CITATION = doi:10.1007/JHEP02(2014)091;%%
  %68 citations counted in INSPIRE as of 29 Aug 2017
  
  
  %\cite{Abada:2014kba}
\bibitem{Abada:2014kba} 
  A.~Abada, M.~E.~Krauss, W.~Porod, F.~Staub, A.~Vicente and C.~Weiland,
  %``Lepton flavor violation in low-scale seesaw models: SUSY and non-SUSY contributions,''
  JHEP {\bf 1411}, 048 (2014)
  doi:10.1007/JHEP11(2014)048
  [arXiv:1408.0138 [hep-ph]].
  %%CITATION = doi:10.1007/JHEP11(2014)048;%%
  %51 citations counted in INSPIRE as of 29 Aug 2017
  
  
  %%%%%%%%%%%%%%%%%%%%%%Theory Study Examples%%%%%%%%%%%%%%%%
  
  %\cite{Antusch:2016vyf}
\bibitem{Antusch:2016vyf} 
  S.~Antusch, E.~Cazzato and O.~Fischer,
  %``Displaced vertex searches for sterile neutrinos at future lepton colliders,''
  JHEP {\bf 1612}, 007 (2016)
  doi:10.1007/JHEP12(2016)007
  [arXiv:1604.02420 [hep-ph]].
  %%CITATION = doi:10.1007/JHEP12(2016)007;%%
  %28 citations counted in INSPIRE as of 28 May 2017
  
  
  %\cite{Helo:2013esa}
\bibitem{Helo:2013esa} 
  J.~C.~Helo, M.~Hirsch and S.~Kovalenko,
  %``Heavy neutrino searches at the LHC with displaced vertices,''
  Phys.\ Rev.\ D {\bf 89}, 073005 (2014)
  Erratum: [Phys.\ Rev.\ D {\bf 93}, no. 9, 099902 (2016)]
  doi:10.1103/PhysRevD.89.073005, 10.1103/PhysRevD.93.099902
  [arXiv:1312.2900 [hep-ph]].
  %%CITATION = doi:10.1103/PhysRevD.89.073005, 10.1103/PhysRevD.93.099902;%%
  %63 citations counted in INSPIRE as of 28 May 2017

  
  %\cite{Dib:2015oka}
\bibitem{Dib:2015oka}
  C.~O.~Dib and C.~S.~Kim,
  %``Discovering sterile Neutrinos ligther than $M_W$ at the LHC,''
  Phys.\ Rev.\ D {\bf 92} (2015) no.9,  093009
  doi:10.1103/PhysRevD.92.093009
  [arXiv:1509.05981 [hep-ph]].
  %%CITATION = doi:10.1103/PhysRevD.92.093009;%%
  %28 citations counted in INSPIRE as of 28 May 2017
  
  %\cite{Mitra:2016kov}
\bibitem{Mitra:2016kov} 
  M.~Mitra, R.~Ruiz, D.~J.~Scott and M.~Spannowsky,
  %``Neutrino Jets from High-Mass $W_R$ Gauge Bosons in TeV-Scale Left-Right Symmetric Models,''
  Phys.\ Rev.\ D {\bf 94}, no. 9, 095016 (2016)
  doi:10.1103/PhysRevD.94.095016
  [arXiv:1607.03504 [hep-ph]].
  %%CITATION = doi:10.1103/PhysRevD.94.095016;%%
  %17 citations counted in INSPIRE as of 28 May 2017
  
  %\cite{Das:2017zjc}
\bibitem{Das:2017zjc} 
  A.~Das, P.~S.~B.~Dev and C.~S.~Kim,
  %``Constraining Sterile Neutrinos from Precision Higgs Data,''
  arXiv:1704.00880 [hep-ph].
  %%CITATION = ARXIV:1704.00880;%%
  
  %\cite{Das:2017rsu}
\bibitem{Das:2017rsu} 
  A.~Das, Y.~Gao and T.~Kamon,
  %``Heavy Neutrino Search via the Higgs boson at the LHC,''
  arXiv:1704.00881 [hep-ph].
  %%CITATION = ARXIV:1704.00881;%%
  
    
  
  %%%%%%%%%%%%%%%% Experimental %%%%%%%%%%%%%%%%%%%%%%%%%%%%%%
  
  %\cite{Khachatryan:2015gha}
\bibitem{Khachatryan:2015gha} 
  V.~Khachatryan {\it et al.} [CMS Collaboration],
  %``Search for heavy Majorana neutrinos in $\mu^\pm \mu^\pm+$ jets events in proton-proton collisions at $\sqrt{s}$ = 8 TeV,''
  Phys.\ Lett.\ B {\bf 748}, 144 (2015)
  doi:10.1016/j.physletb.2015.06.070
  [arXiv:1501.05566 [hep-ex]].
  %%CITATION = doi:10.1016/j.physletb.2015.06.070;%%
  %64 citations counted in INSPIRE as of 18 May 2017
  
  %\cite{Khachatryan:2016olu}
\bibitem{Khachatryan:2016olu} 
  V.~Khachatryan {\it et al.} [CMS Collaboration],
  %``Search for heavy Majorana neutrinos in e$^{±}$e$^{±}$+ jets and e$^{±}$ $\mu^{±}$+ jets events in proton-proton collisions at $ \sqrt{s}=8 $ TeV,''
  JHEP {\bf 1604}, 169 (2016)
  doi:10.1007/JHEP04(2016)169
  [arXiv:1603.02248 [hep-ex]].
  %%CITATION = doi:10.1007/JHEP04(2016)169;%%
  %21 citations counted in INSPIRE as of 18 May 2017

  
  %\cite{Aad:2015xaa}
\bibitem{Aad:2015xaa} 
  G.~Aad {\it et al.} [ATLAS Collaboration],
  %``Search for heavy Majorana neutrinos with the ATLAS detector in pp collisions at $ \sqrt{s}=8 $ TeV,''
  JHEP {\bf 1507}, 162 (2015)
  doi:10.1007/JHEP07(2015)162
  [arXiv:1506.06020 [hep-ex]].
  %%CITATION = doi:10.1007/JHEP07(2015)162;%%
  %76 citations counted in INSPIRE as of 18 May 2017

  
%\cite{Sirunyan:2017xnz}
\bibitem{Sirunyan:2017xnz} 
  A.~M.~Sirunyan {\it et al.} [CMS Collaboration],
  %``Search for a heavy composite Majorana neutrino in the final state with two leptons and two quarks at sqrt(s) = 13 TeV,''
  arXiv:1706.08578 [hep-ex].
  %%CITATION = ARXIV:1706.08578;%%
  
  %%%%%%%%%%%%%% First paper on the topic%%%%%%%%%%%%%%%%%%%%%
 
 %\cite{Izaguirre:2015pga}
\bibitem{Izaguirre:2015pga}
  E.~Izaguirre and B.~Shuve,
  %``Multilepton and Lepton Jet Probes of Sub-Weak-Scale Right-Handed Neutrinos,''
  Phys.\ Rev.\ D {\bf 91} (2015) no.9,  093010
  doi:10.1103/PhysRevD.91.093010
  [arXiv:1504.02470 [hep-ph]].
  %%CITATION = doi:10.1103/PhysRevD.91.093010;%%
  %26 citations counted in INSPIRE as of 07 Dec 2016
  
  
  
  %%%%%%%%%%%%%%100 TeV collider %%%%%%%%%%%%%%%%%%%
  
  %\cite{Golling:2016gvc}
\bibitem{Golling:2016gvc} 
  T.~Golling {\it et al.},
  %``Physics at a 100 TeV pp collider: beyond the Standard Model phenomena,''
  %Submitted to: Phys.Rept.
  [arXiv:1606.00947 [hep-ph]].
  %%CITATION = ARXIV:1606.00947;%%
  %39 citations counted in INSPIRE as of 28 May 2017
  
  
  
%\cite{CEPC-SPPCStudyGroup:2015csa}
\bibitem{CEPC-SPPCStudyGroup:2015csa}
  CEPC-SPPC Study Group,
  %``CEPC-SPPC Preliminary Conceptual Design Report. 1. Physics and Detector,''
  IHEP-CEPC-DR-2015-01, IHEP-TH-2015-01, IHEP-EP-2015-01.
  %%CITATION = IHEP-CEPC-DR-2015-01, IHEP-TH-2015-01, IHEP-EP-2015-01;%%
  %22 citations counted in INSPIRE as of 23 May 2017

   
   
   %%%%%%%%%% General Lepton jets %%%%%%%%%%%%%%

  %\cite{Aad:2014yea}
\bibitem{Aad:2014yea} 
  G.~Aad {\it et al.} [ATLAS Collaboration],
  %``Search for long-lived neutral particles decaying into lepton jets in proton-proton collisions at $ \sqrt{s}=8 $ TeV with the ATLAS detector,''
  JHEP {\bf 1411}, 088 (2014)
  doi:10.1007/JHEP11(2014)088
  [arXiv:1409.0746 [hep-ex]].
  %%CITATION = doi:10.1007/JHEP11(2014)088;%%
  %72 citations counted in INSPIRE as of 18 May 2017
  

  
  %%%%%%%%%%% EWPD constraint %%%%%%%%%%%%%%%%%
  
  \bibitem{delAguila:2008pw}
F.~del Aguila, J.~de~Blas, and M.~Perez-Victoria,
\newblock Phys.Rev. {\bf D78}, 013010 (2008), 0803.4008.

\bibitem{Akhmedov:2013hec}
E.~Akhmedov, A.~Kartavtsev, M.~Lindner, L.~Michaels, and J.~Smirnov,
\newblock JHEP {\bf 1305}, 081 (2013), 1302.1872.

\bibitem{Basso:2013jka}
L.~Basso, O.~Fischer, and J.~J. van~der Bij,
\newblock Europhys.Lett. {\bf 105}, 11001 (2014), 1310.2057.

\bibitem{Blas:2013ana}
J.~de~Blas,
\newblock EPJ Web Conf. {\bf 60}, 19008 (2013), 1307.6173.

\bibitem{Antusch:2015mia}
S.~Antusch and O.~Fischer,
\newblock (2015), 1502.05915.


%%%%%%%%%%%%%%Constraints from Z-decays%%%%%%%%%%%%%


\bibitem{Adriani:1992pq}
L3 Collaboration, O.~Adriani {\em et~al.},
\newblock Phys.Lett. {\bf B295}, 371 (1992).

\bibitem{Abreu:1996pa}
DELPHI Collaboration, P.~Abreu {\em et~al.},
\newblock Z.Phys. {\bf C74}, 57 (1997).

%\cite{Shevchenko:1993fe}
\bibitem{Shevchenko:1993fe} 
  S.~Shevchenko,
  %``Search for isosinglet neutral heavy lepton with the L3 detector at LEP,''
  CALT-68-1939.
  %%CITATION = CALT-68-1939;%%
  
   %%%%%%%%%%%%%%%%%%% PDG %%%%%%%%%%%%%%%%%%%%
  
  %\cite{Olive:2016xmw}
\bibitem{Olive:2016xmw}
  C.~Patrignani {\it et al.} [Particle Data Group],
  %``Review of Particle Physics,''
  Chin.\ Phys.\ C {\bf 40} (2016) no.10,  100001.
  doi:10.1088/1674-1137/40/10/100001
  %%CITATION = doi:10.1088/1674-1137/40/10/100001;%%
  %804 citations counted in INSPIRE as of 09 May 2017

       
  
  %%%%%%%%%%%%%%% Neutrino mass scales %%%%%%%%%%%%%%%%

  
   % \cite{troitsk}
 \bibitem{troitsk}
  V.~N.~Aseev {\it et al.} [Troitsk Collaboration],
  %``An upper limit on electron antineutrino mass from Troitsk experiment,''
  Phys.\ Rev.\ D {\bf 84}, 112003 (2011) doi:10.1103/PhysRevD.84.112003
  [arXiv:1108.5034 [hep-ex]].
  %%CITATION = doi:10.1103/PhysRevD.84.112003;%%
  %236 citations counted in INSPIRE as of 09 Jun 2017


  % \cite{mainz}
\bibitem{mainz} 
  C.~Kraus {\it et al.},
  %``Final results from phase II of the Mainz neutrino mass search in tritium beta decay,''
  Eur.\ Phys.\ J.\ C {\bf 40}, 447 (2005)
  doi:10.1140/epjc/s2005-02139-7
  [hep-ex/0412056].
  %%CITATION = doi:10.1140/epjc/s2005-02139-7;%%
  %601 citations counted in INSPIRE as of 09 Jun 2017

 
  %\cite{Reid:2009nq}
\bibitem{Reid:2009nq}
  B.~A.~Reid, L.~Verde, R.~Jimenez, O.~Mena,
  %``Robust Neutrino Constraints by Combining Low Redshift Observations with the CMB,''
  JCAP {\bf 1001}, 003 (2010).
  [arXiv:0910.0008 [astro-ph.CO]].
  
  %\cite{GonzalezGarcia:2010un}
\bibitem{GonzalezGarcia:2010un}
  M.~C.~Gonzalez-Garcia, M.~Maltoni, J.~Salvado,
  %``Robust Cosmological Bounds on Neutrinos and their Combination with Oscillation Results,''
  JHEP {\bf 1008}, 117 (2010).
  [arXiv:1006.3795 [hep-ph]].
  
  
  %\cite{Ade:2015xua}
\bibitem{Ade:2015xua} 
  P.~A.~R.~Ade {\it et al.} [Planck Collaboration],
  %``Planck 2015 results. XIII. Cosmological parameters,''
  Astron.\ Astrophys.\  {\bf 594}, A13 (2016)
  doi:10.1051/0004-6361/201525830
  [arXiv:1502.01589 [astro-ph.CO]].
  %%CITATION = doi:10.1051/0004-6361/201525830;%%
  %3449 citations counted in INSPIRE as of 09 Jun 2017
  
  %\cite{KATRIN}
\bibitem{KATRIN}
% KATRIN: Direct Measurement of a sub-eV Neutrino Mass
G.~Drexlin et al,
Nucl. Phys. B (Proc. Suppl.) 145 (2005) 263-267

  %%%%%%%%% GUTs %%%%%%%%%%

%\cite{Langacker:1980js}
\bibitem{Langacker:1980js} 
  P.~Langacker,
  %``Grand Unified Theories and Proton Decay,''
  Phys.\ Rept.\  {\bf 72}, 185 (1981).
  doi:10.1016/0370-1573(81)90059-4
  %%CITATION = doi:10.1016/0370-1573(81)90059-4;%%
  %1299 citations counted in INSPIRE as of 09 Jun 2017
  
  
  %\cite{Pati:1974yy}
\bibitem{Pati:1974yy} 
  J.~C.~Pati and A.~Salam,
  %``Lepton Number as the Fourth Color,''
  Phys.\ Rev.\ D {\bf 10}, 275 (1974)
  Erratum: [Phys.\ Rev.\ D {\bf 11}, 703 (1975)].
  doi:10.1103/PhysRevD.10.275, 10.1103/PhysRevD.11.703.2
  %%CITATION = doi:10.1103/PhysRevD.10.275, 10.1103/PhysRevD.11.703.2;%%
  %4216 citations counted in INSPIRE as of 09 Jun 2017


     
  %%%% Extended see-saw models %%% 
  
%\cite{Wyler:1982dd}
\bibitem{Wyler:1982dd} 
  D.~Wyler and L.~Wolfenstein,
  %``Massless Neutrinos in Left-Right Symmetric Models,''
  Nucl.\ Phys.\ B {\bf 218}, 205 (1983).
  %%CITATION = NUPHA,B218,205;%%
  %250 citations counted in INSPIRE as of 20 mar 2015
  \\[-10pt]
  
  %\cite{Mohapatra:1986aw}
\bibitem{Mohapatra:1986aw}
  R.~N.~Mohapatra,
  %``Mechanism for Understanding Small Neutrino Mass in Superstring Theories,''
  Phys.\ Rev.\ Lett.\  {\bf 56} (1986) 561.
  doi:10.1103/PhysRevLett.56.561
  %%CITATION = doi:10.1103/PhysRevLett.56.561;%%
  %338 citations counted in INSPIRE as of 11 May 2017

%\cite{Mohapatra:1986bd}
\bibitem{Mohapatra:1986bd} 
  R.~N.~Mohapatra and J.~W.~F.~Valle,
  %``Neutrino Mass and Baryon Number Nonconservation in Superstring Models,''
  Phys.\ Rev.\ D {\bf 34}, 1642 (1986).
  %%CITATION = PHRVA,D34,1642;%%
  %620 citations counted in INSPIRE as of 20 mar 2015
  \\[-10pt]
  
  %\cite{Langacker:1998ut}
\bibitem{Langacker:1998ut}
  P.~Langacker,
  %``A Mechanism for ordinary sterile neutrino mixing,''
  Phys.\ Rev.\ D {\bf 58} (1998) 093017
  doi:10.1103/PhysRevD.58.093017
  [hep-ph/9805281].
  %%CITATION = doi:10.1103/PhysRevD.58.093017;%%
  %112 citations counted in INSPIRE as of 11 May 2017
  
  %\cite{Chacko:2003dt}
\bibitem{Chacko:2003dt}
  Z.~Chacko, L.~J.~Hall, T.~Okui and S.~J.~Oliver,
  %``CMB signals of neutrino mass generation,''
  Phys.\ Rev.\ D {\bf 70} (2004) 085008
  doi:10.1103/PhysRevD.70.085008
  [hep-ph/0312267].
  %%CITATION = doi:10.1103/PhysRevD.70.085008;%%
  %61 citations counted in INSPIRE as of 11 May 2017
  
  %\cite{Gherghetta:2003hf}
\bibitem{Gherghetta:2003hf}
  T.~Gherghetta,
  %``Dirac neutrino masses with Planck scale lepton number violation,''
  Phys.\ Rev.\ Lett.\  {\bf 92} (2004) 161601
  doi:10.1103/PhysRevLett.92.161601
  [hep-ph/0312392].
  %%CITATION = doi:10.1103/PhysRevLett.92.161601;%%
  %71 citations counted in INSPIRE as of 11 May 2017
 
%\cite{Malinsky:2005bi}
\bibitem{Malinsky:2005bi} 
  M.~Malinsky, J.~C.~Romao and J.~W.~F.~Valle,
  %``Novel supersymmetric SO(10) seesaw mechanism,''
  Phys.\ Rev.\ Lett.\  {\bf 95}, 161801 (2005)
  [hep-ph/0506296].
  %%CITATION = HEP-PH/0506296;%%
  %119 citations counted in INSPIRE as of 20 mar 2015
  \\[-10pt]
  
%\cite{Shaposhnikov:2006nn}
\bibitem{Shaposhnikov:2006nn} 
  M.~Shaposhnikov,
  %``A Possible symmetry of the nuMSM,''
  Nucl.\ Phys.\ B {\bf 763}, 49 (2007)
  [hep-ph/0605047].
  %%CITATION = HEP-PH/0605047;%%
  %107 citations counted in INSPIRE as of 20 mar 2015
  \\[-10pt]
  
   %\cite{deGouvea:2006gz}
\bibitem{deGouvea:2006gz}
  A.~de Gouvea, J.~Jenkins and N.~Vasudevan,
  %``Neutrino Phenomenology of Very Low-Energy Seesaws,''
  Phys.\ Rev.\ D {\bf 75} (2007) 013003
  doi:10.1103/PhysRevD.75.013003
  [hep-ph/0608147].
  %%CITATION = doi:10.1103/PhysRevD.75.013003;%%
  %98 citations counted in INSPIRE as of 11 May 2017
  
%\cite{Kersten:2007vk}
\bibitem{Kersten:2007vk} 
  J.~Kersten and A.~Y.~Smirnov,
  %``Right-Handed Neutrinos at CERN LHC and the Mechanism of Neutrino Mass Generation,''
  Phys.\ Rev.\ D {\bf 76}, 073005 (2007)
  [arXiv:0705.3221 [hep-ph]].
  %%CITATION = ARXIV:0705.3221;%%
  %206 citations counted in INSPIRE as of 20 mar 2015
  \\[-10pt]

%\cite{Gavela:2009cd}
\bibitem{Gavela:2009cd} 
  M.~B.~Gavela, T.~Hambye, D.~Hernandez and P.~Hernandez,
  %``Minimal Flavour Seesaw Models,''
  JHEP {\bf 0909}, 038 (2009)
  [arXiv:0906.1461 [hep-ph]].
  %%CITATION = ARXIV:0906.1461;%%
  %98 citations counted in INSPIRE as of 20 mar 2015
  
  
  
   %\cite{deGouvea:2011zz}
\bibitem{deGouvea:2011zz}
  A.~de Gouvea and W.~C.~Huang,
  %``Constraining the (Low-Energy) Type-I Seesaw,''
  Phys.\ Rev.\ D {\bf 85} (2012) 053006
  doi:10.1103/PhysRevD.85.053006
  [arXiv:1110.6122 [hep-ph]].
  %%CITATION = doi:10.1103/PhysRevD.85.053006;%%
  %23 citations counted in INSPIRE as of 11 May 2017
  
  %%%%%%%%%%%%%% N production %%%%%%%%%%%%%%%
  
  %\cite{Dev:2013wba}
\bibitem{Dev:2013wba} 
  P.~S.~B.~Dev, A.~Pilaftsis and U.~k.~Yang,
  %``New Production Mechanism for Heavy Neutrinos at the LHC,''
  Phys.\ Rev.\ Lett.\  {\bf 112}, no. 8, 081801 (2014)
  doi:10.1103/PhysRevLett.112.081801
  [arXiv:1308.2209 [hep-ph]].
  %%CITATION = doi:10.1103/PhysRevLett.112.081801;%%
  %81 citations counted in INSPIRE as of 09 Jun 2017
  
  %%%%%%%%% References added for production modes %%%%%%%%
  
  %\cite{Alva:2014gxa}
\bibitem{Alva:2014gxa} 
  D.~Alva, T.~Han and R.~Ruiz,
  %``Heavy Majorana neutrinos from $W\gamma$ fusion at hadron colliders,''
  JHEP {\bf 1502}, 072 (2015)
  doi:10.1007/JHEP02(2015)072
  [arXiv:1411.7305 [hep-ph]].
  %%CITATION = doi:10.1007/JHEP02(2015)072;%%
  %42 citations counted in INSPIRE as of 29 Aug 2017
  
  %\cite{Hessler:2014ssa}
\bibitem{Hessler:2014ssa} 
  A.~G.~Hessler, A.~Ibarra, E.~Molinaro and S.~Vogl,
  %``Impact of the Higgs boson on the production of exotic particles at the LHC,''
  Phys.\ Rev.\ D {\bf 91}, no. 11, 115004 (2015)
  doi:10.1103/PhysRevD.91.115004
  [arXiv:1408.0983 [hep-ph]].
  %%CITATION = doi:10.1103/PhysRevD.91.115004;%%
  %12 citations counted in INSPIRE as of 29 Aug 2017
  
  %\cite{Degrande:2016aje}
\bibitem{Degrande:2016aje} 
  C.~Degrande, O.~Mattelaer, R.~Ruiz and J.~Turner,
  %``Fully-Automated Precision Predictions for Heavy Neutrino Production Mechanisms at Hadron Colliders,''
  Phys.\ Rev.\ D {\bf 94}, no. 5, 053002 (2016)
  doi:10.1103/PhysRevD.94.053002
  [arXiv:1602.06957 [hep-ph]].
  %%CITATION = doi:10.1103/PhysRevD.94.053002;%%
  %23 citations counted in INSPIRE as of 29 Aug 2017

  
  %\cite{Ruiz:2017yyf}
\bibitem{Ruiz:2017yyf} 
  R.~Ruiz, M.~Spannowsky and P.~Waite,
  %``Heavy Neutrino Production from Threshold Resummed Gluon Fusion,''
  arXiv:1706.02298 [hep-ph].
  %%CITATION = ARXIV:1706.02298;%%
  
  
    
  %%%%%%% Production, BR and partial width theory %%%%%%%%%
  
  
  %\cite{Helo:2010cw}
\bibitem{Helo:2010cw}
  J.~C.~Helo, S.~Kovalenko and I.~Schmidt,
  %``Sterile neutrinos in lepton number and lepton flavor violating decays,''
  Nucl.\ Phys.\ B {\bf 853} (2011) 80
  doi:10.1016/j.nuclphysb.2011.07.020
  [arXiv:1005.1607 [hep-ph]].
  %%CITATION = doi:10.1016/j.nuclphysb.2011.07.020;%%
  %41 citations counted in INSPIRE as of 14 May 2017


  %%%%%%%%%%%%%%%%%%% ATLAS/CMS ML searches %%%%%%%%%%%%%%%%%%%%
  %\cite{CMS:2017wua}
\bibitem{CMS:2017wua} 
  CMS Collaboration [CMS Collaboration],
  %``Search for evidence of Type-III seesaw mechanism in multilepton final states in pp collisions at $\sqrt{s} = 13~\mathrm{TeV}$,''
  CMS-PAS-EXO-17-006.
  %%CITATION = CMS-PAS-EXO-17-006;%%
  %1 citations counted in INSPIRE as of 09 Jun 2017

%\cite{Aad:2014hja}
\bibitem{Aad:2014hja} 
  G.~Aad {\it et al.} [ATLAS Collaboration],
  %``Search for new phenomena in events with three or more charged leptons in $pp$ collisions at $\sqrt{s}=8$ TeV with the ATLAS detector,''
  JHEP {\bf 1508}, 138 (2015)
  doi:10.1007/JHEP08(2015)138
  [arXiv:1411.2921 [hep-ex]].
  %%CITATION = doi:10.1007/JHEP08(2015)138;%%
  %21 citations counted in INSPIRE as of 09 Jun 2017
  
  %%%%%%%%%%%%%%%%%%  ATLAS and CMS Lep-jet searches %%%%%%%%%%%%%%%%%%%%
  %\cite{ATLAS:2016jza}
\bibitem{ATLAS:2016jza}
  The ATLAS Collaboration,
  %Search for long-lived neutral particles decaying into  displaced lepton jets in proton--proton collisions at  $\sqrt{s}$ = 13 TeV with the ATLAS detector
ATLAS-CONF-2016-042.
 "%%CITATION = ATLAS-CONF-2016-042;%%"

%\cite{CMS:2014hka}
\bibitem{CMS:2014hka} 
  The CMS Collaboration,
  %``Search for long-lived particles that decay into final states containing two electrons or two muons in proton-proton collisions at $\sqrt{s} =$ 8 TeV,''
  %  Phys.\ Rev.\ D {\bf 91}, no. 5, 052012 (2015)
    Phys.\ Rev.\ D {\bf 91} (2015) 052012
  doi:10.1103/PhysRevD.91.052012
  [arXiv:1411.6977 [hep-ex]].
  %%CITATION = doi:10.1103/PhysRevD.91.052012;%%
  %58 citations counted in INSPIRE as of 08 Jun 2017
  
  
  %%%%%%%%%%%%%%%%%%%%  LBNE/DUNE & SHiP %%%%%%%%%%%%%%%%%%%%%%%%
  
  %\cite{Acciarri:2016crz}
\bibitem{Acciarri:2016crz} 
  R.~Acciarri {\it et al.} [DUNE Collaboration],
  %``Long-Baseline Neutrino Facility (LBNF) and Deep Underground Neutrino Experiment (DUNE) : Volume 1: The LBNF and DUNE Projects,''
  arXiv:1601.05471 [physics.ins-det].
  %%CITATION = ARXIV:1601.05471;%%
  %75 citations counted in INSPIRE as of 05 Jun 2017
  
  %\cite{Acciarri:2015uup}
\bibitem{Acciarri:2015uup} 
  R.~Acciarri {\it et al.} [DUNE Collaboration],
  %``Long-Baseline Neutrino Facility (LBNF) and Deep Underground Neutrino Experiment (DUNE) : Volume 2: The Physics Program for DUNE at LBNF,''
  arXiv:1512.06148 [physics.ins-det].
  %%CITATION = ARXIV:1512.06148;%%
  %156 citations counted in INSPIRE as of 05 Jun 2017
  
  %\cite{Alekhin:2015byh}
\bibitem{Alekhin:2015byh} 
  S.~Alekhin {\it et al.},
  %``A facility to Search for Hidden Particles at the CERN SPS: the SHiP physics case,''
  Rept.\ Prog.\ Phys.\  {\bf 79}, no. 12, 124201 (2016)
  doi:10.1088/0034-4885/79/12/124201
  [arXiv:1504.04855 [hep-ph]].
  %%CITATION = doi:10.1088/0034-4885/79/12/124201;%%
  %167 citations counted in INSPIRE as of 05 Jun 2017
  
  %\cite{DeLellis:2017rfg}
\bibitem{DeLellis:2017rfg} 
  G.~De Lellis [SHiP Collaboration],
  %``The SHiP experiment at CERN,''
  Nucl.\ Part.\ Phys.\ Proc.\  {\bf 285-286}, 126.
  doi:10.1016/j.nuclphysbps.2017.03.023
  %%CITATION = doi:10.1016/j.nuclphysbps.2017.03.023;%%


%\cite{Gray:2011us}
\bibitem{Gray:2011us} 
  R.~C.~Gray, C.~Kilic, M.~Park, S.~Somalwar and S.~Thomas,
  %``Backgrounds To Higgs Boson Searches from $W \gamma^* -> l \nu l (l)$ Asymmetric Internal Conversion,''
  arXiv:1110.1368 [hep-ph].
  %%CITATION = ARXIV:1110.1368;%%
  %46 citations counted in INSPIRE as of 09 Jun 2017

  
   
  %%%%%%%%%%%%%%%%%%%%%%%%%%%%%%MC generation%%%%%%%%%%%%%%%%%%%%%
  
     %\cite{Alwall:2014hca}
\bibitem{Alwall:2014hca} 
  J.~Alwall {\it et al.},
  %``The automated computation of tree-level and next-to-leading order differential cross sections, and their matching to parton shower simulations,''
  JHEP {\bf 1407}, 079 (2014)
  doi:10.1007/JHEP07(2014)079
  [arXiv:1405.0301 [hep-ph]].
  %%CITATION = doi:10.1007/JHEP07(2014)079;%%
  %1677 citations counted in INSPIRE as of 08 Mar 2017 
  
  %\cite{Sjostrand:2014zea}
\bibitem{Sjostrand:2014zea}
  T.~Sj\"{o}strand {\it et al.},
  %``An Introduction to PYTHIA 8.2,''
  Comput.\ Phys.\ Commun.\  {\bf 191} (2015) 159
  doi:10.1016/j.cpc.2015.01.024
  [arXiv:1410.3012 [hep-ph]].
  %%CITATION = doi:10.1016/j.cpc.2015.01.024;%%
  %180 citations counted in INSPIRE as of 25 Mar 2016
      
   %\cite{deFavereau:2013fsa}
\bibitem{deFavereau:2013fsa}
  J.~de Favereau {\it et al.} [DELPHES 3 Collaboration],
  %``DELPHES 3, A modular framework for fast simulation of a generic collider experiment,''
  JHEP {\bf 1402} (2014) 057
  doi:10.1007/JHEP02(2014)057
  [arXiv:1307.6346 [hep-ex]].
  %%CITATION = doi:10.1007/JHEP02(2014)057;%%
  %637 citations counted in INSPIRE as of 08 Mar 2017

   %\cite{CMS:2016btag}
\bibitem{CMS:2016btag} 
  The CMS Collaboration,
  %"{Identification of b quark jets at the CMS Experiment in the LHC Run 2}" 
  CMS-PAS-BTV-15-001 (2016)

  %%CITATION = CMS-PAS-BTV-15-001 (2016)%%
  
 
%%%%%%%%%%%%%%%% CLs method%%%%%%%%%%%%%%%%%%

%\cite{Junk:1999kv}
\bibitem{Junk:1999kv}
  T.~Junk,
  %``Confidence level computation for combining searches with small statistics,''
  Nucl.\ Instrum.\ Meth.\ A {\bf 434} (1999) 435
  doi:10.1016/S0168-9002(99)00498-2
  [hep-ex/9902006].
  %%CITATION = doi:10.1016/S0168-9002(99)00498-2;%%
  %1289 citations counted in INSPIRE as of 19 Mar 2017

\bibitem{CLsConf}
A. L. Read, “Modified frequentist analysis of search results (the CLs method) in
“Workshop on Confidence Limits”, Eds. F. James, L. Lyons, and Y. Perrin”,. p. 81.

\bibitem{Read:722145}
A~L Read.
\newblock {Presentation of search results: the CL$_{s}$ technique}.
\newblock {\em J. Phys. G}, 28(10):2693--704, 2002.

\bibitem{CMS-NOTE-2011-005}
{Procedure for the LHC Higgs boson search combination in Summer 2011}.
\newblock Technical Report CMS-NOTE-2011-005. ATL-PHYS-PUB-2011-11, CERN,
  Geneva, Aug 2011.  
  
    %%%%%%%%%%%%%%%%%%%%% LHCb analysis %%%%%%%%%%%%%%%%%  
    
    %\cite{Antusch:2017hhu}
\bibitem{Antusch:2017hhu} 
  S.~Antusch, E.~Cazzato and O.~Fischer,
  %``Sterile neutrino searches via displaced vertices at LHCb,''
  arXiv:1706.05990 [hep-ph].
  %%CITATION = ARXIV:1706.05990;%%
  
  %\cite{Aaij:2016xmb}
\bibitem{Aaij:2016xmb} 
  R.~Aaij {\it et al.} [LHCb Collaboration],
  %``Search for massive long-lived particles decaying semileptonically in the LHCb detector,''
  Eur.\ Phys.\ J.\ C {\bf 77}, no. 4, 224 (2017)
  doi:10.1140/epjc/s10052-017-4744-6
  [arXiv:1612.00945 [hep-ex]].
  %%CITATION = doi:10.1140/epjc/s10052-017-4744-6;%%
  %4 citations counted in INSPIRE as of 20 Jun 2017
  
  
    
\end{thebibliography}\endgroup
\bibliographystyle{jhep}
%------------------------------------------------------------------------------------------------------------------ 

\end{document}
%------------------------------------------------------------------------------------------------------------------